\documentclass[twocolumn]{aastex631}
\usepackage{amsmath}
\usepackage{graphicx}

\usepackage{natbib}
\usepackage[scaled]{helvet}
\usepackage{epsfig}
\usepackage{url}
\bibpunct{(}{)}{;}{a}{}{,}
\newcommand{\kms}{\,km\,s$^{-1}$}
\usepackage{textcomp}
\usepackage{mathpazo}
\usepackage{xspace}
\usepackage{hyperref}
\usepackage{apjfonts}
\usepackage{mathrsfs}
\usepackage{verbatim}

\usepackage{color}
\usepackage[normalem]{ulem}

\newcommand{\bjdtdb}{\ensuremath{\rm {BJD_{TDB}}}}

\newcommand{\msun}{\ensuremath{\,M_\Sun}}
\newcommand{\rsun}{\ensuremath{\,R_\Sun}}

\newcommand{\mj}{\ensuremath{\,M_{\rm J}}}
\newcommand{\rj}{\ensuremath{\,R_{\rm J}}}

\newcommand{\fave}{\langle F \rangle}
\newcommand{\fluxcgs}{10$^9$ erg s$^{-1}$ cm$^{-2}$}

\newcommand{\vsini}{\ensuremath{v\sin{i_*}}}

\newcommand{\thisstar}{HIP 33609\xspace}
\newcommand{\thisplanet}{HIP 33609 b\xspace}
\newcommand{\mstar}{\ensuremath{M_{*}}}
\newcommand{\rstar}{\ensuremath{R_{*}}}

\newcommand{\be}{\begin{equation}}
\newcommand{\ee}{\end{equation}}

\newcommand{\association}{MELANGE-6}
\newcommand{\tess}{{\it TESS}}

\newcommand{\rplanet}{{R$_{\rm b}$ = 1.580$_{-0.070}^{+0.074}$ R$_{\rm J}$}}
\newcommand{\mplanet}{{M$_{\rm b}$ = 68.0$_{-7.1}^{+7.4}$ M$_{\rm J}$}}
\newcommand{\starmass}{{\mstar = 2.383$_{-0.095}^{+0.10}$ M$_{\odot}$}}
\newcommand{\starrad}{{\rstar = 1.863$_{-0.082}^{+0.087}$ R$_{\odot}$}}
\newcommand{\ecc}{{e = 0.560$_{-0.031}^{+0.029}$}}
\newcommand{\period}{p = 39.47 days}
\newcommand{\teffstar}{{T$_{\rm eff}$ = 10,400$_{-660}^{+800}$ K}}

\begin{document}

\title{HIP 33609 b: An Eccentric Brown Dwarf Transiting a V=7.3 Rapidly Rotating B-Star}

\newcommand{\cfa}{Center for Astrophysics \textbar \ Harvard \& Smithsonian, 60 Garden St, Cambridge, MA 02138, USA}
\newcommand{\msu}{Center for Data Intensive and Time Domain Astronomy, Department of Physics and Astronomy, Michigan State University, East Lansing, MI 48824, USA}
\newcommand{\umich}{Astronomy Department, University of Michigan, 1085 S University Avenue, Ann Arbor, MI 48109, USA}
\newcommand{\utaustin}{Department of Astronomy, The University of Texas at Austin, Austin, TX 78712, USA}
\newcommand{\MIT}{Department of Physics and Kavli Institute for Astrophysics and Space Research, Massachusetts Institute of Technology, Cambridge, MA 02139, USA}
\newcommand{\MITEPS}{Department of Earth, Atmospheric and Planetary Sciences, Massachusetts Institute of Technology,  Cambridge,  MA 02139, USA}
\newcommand{\uflorida}{Department of Astronomy, University of Florida, 211 Bryant Space Science Center, Gainesville, FL, 32611, USA}
\newcommand{\riverside}{Department of Earth Sciences, University of California, Riverside, CA 92521, USA}
\newcommand{\usq}{University of Southern Queensland, West St, Darling Heights QLD 4350, Australia}
\newcommand{\ames}{NASA Ames Research Center, Moffett Field, CA, 94035, USA}
\newcommand{\geneva}{Geneva Observatory, University of Geneva, Chemin des Mailettes 51, 1290 Versoix, Switzerland}
\newcommand{\uw}{Astronomy Department, University of Washington, Seattle, WA 98195 USA}
\newcommand{\warwick}{Deptartment of Physics, University of Warwick, Gibbet Hill Road, Coventry CV4 7AL, UK}
\newcommand{\warwickceh}{Centre for Exoplanets and Habitability, University of Warwick, Gibbet Hill Road, Coventry CV4 7AL, UK}
\newcommand{\princeton}{Department of Astrophysical Sciences, Princeton University, 4 Ivy Lane, Princeton, NJ, 08544, USA}
\newcommand{\liege}{Space Sciences, Technologies and Astrophysics Research (STAR) Institute, Universit\'e de Li\`ege, 19C All\'ee du 6 Ao\^ut, 4000 Li\`ege, Belgium}
\newcommand{\vanderbilt}{Department of Physics and Astronomy, Vanderbilt University, Nashville, TN 37235, USA}
\newcommand{\fisk}{Department of Physics, Fisk University, 1000 17th Avenue North, Nashville, TN 37208, USA}
\newcommand{\columbia}{Department of Astronomy, Columbia University, 550 West 120th Street, New York, NY 10027, USA}
\newcommand{\toronto}{Dunlap Institute for Astronomy and Astrophysics, University of Toronto, Ontario M5S 3H4, Canada}
\newcommand{\unc}{Department of Physics and Astronomy, University of North Carolina at Chapel Hill, Chapel Hill, NC 27599, USA}
\newcommand{\iac}{Instituto de Astrof\'isica de Canarias (IAC), E-38205 La Laguna, Tenerife, Spain}
\newcommand{\lalaguna}{Departamento de Astrof\'isica, Universidad de La Laguna (ULL), E-38206 La Laguna, Tenerife, Spain}
\newcommand{\louisville}{Department of Physics and Astronomy, University of Louisville, Louisville, KY 40292, USA}
\newcommand{\aavso}{American Association of Variable Star Observers, 49 Bay State Road, Cambridge, MA 02138, USA}
\newcommand{\utokyo}{The University of Tokyo, 7-3-1 Hongo, Bunky\={o}, Tokyo 113-8654, Japan}
\newcommand{\naoj}{National Astronomical Observatory of Japan, 2-21-1 Osawa, Mitaka, Tokyo 181-8588, Japan}
\newcommand{\jstpresto}{JST, PRESTO, 7-3-1 Hongo, Bunkyo-ku, Tokyo 113-0033, Japan}
\newcommand{\astrobiojapan}{Astrobiology Center, 2-21-1 Osawa, Mitaka, Tokyo 181-8588, Japan}
\newcommand{\ctio}{Cerro Tololo Inter-American Observatory, Casilla 603, La Serena, Chile}
\newcommand{\nexsci}{Caltech IPAC -- NASA Exoplanet Science Institute 1200 E. California Ave, Pasadena, CA 91125, USA}
\newcommand{\ucsc}{Department of Astronomy and Astrophysics, University of
California, Santa Cruz, CA 95064, USA}
\newcommand{\gsfc}{Exoplanets and Stellar Astrophysics Laboratory, Code 667, NASA Goddard Space Flight Center, Greenbelt, MD 20771, USA}
\newcommand{\sgtinc}{SGT, Inc./NASA AMES Research Center, Mailstop 269-3, Bldg T35C, P.O. Box 1, Moffett Field, CA 94035, USA}
\newcommand{\chile}{Center of Astro-Engineering UC, Pontificia Universidad Cat\'olica de Chile, Av. Vicu\~{n}a Mackenna 4860, 7820436 Macul, Santiago, Chile}
\newcommand{\Pontificia}{Instituto de Astrof\'isica, Pontificia Universidad Cat\'olica de Chile, Av.\ Vicu\~na Mackenna 4860, Macul, Santiago, Chile}
\newcommand{\Millennium}{Millennium Institute for Astrophysics, Chile}
\newcommand{\maxplank}{Max-Planck-Institut f\"ur Astronomie, K\"onigstuhl 17, Heidelberg 69117, Germany}
\newcommand{\utdallas}{Department of Physics, The University of Texas at Dallas, 800 West
Campbell Road, Richardson, TX 75080-3021 USA}
\newcommand{\MauryLewin}{Maury Lewin Astronomical Observatory, Glendora, CA 91741, USA}
\newcommand{\umbc}{University of Maryland, Baltimore County, 1000 Hilltop Circle, Baltimore, MD 21250, USA}
\newcommand{\osu}{Department of Astronomy, The Ohio State University, 140 West 18th Avenue, Columbus, OH 43210, USA}
\newcommand{\MITAA}{Department of Aeronautics and Astronautics, MIT, 77 Massachusetts Avenue, Cambridge, MA 02139, USA}
\newcommand{\openu}{School of Physical Sciences, The Open University, Milton Keynes MK7 6AA, UK}
\newcommand{\swarthmore}{Department of Physics and Astronomy, Swarthmore College, Swarthmore, PA 19081, USA}
\newcommand{\seti}{SETI Institute, Mountain View, CA 94043, USA}
\newcommand{\lehigh}{Department of Physics, Lehigh University, 16 Memorial Drive East, Bethlehem, PA 18015, USA}
\newcommand{\utah}{Department of Physics and Astronomy, University of Utah, 115 South 1400 East, Salt Lake City, UT 84112, USA}
\newcommand{\USNA}{Department of Physics, United States Naval Academy, 572C Holloway Rd., Annapolis, MD 21402, USA}
\newcommand{\DTM}{Department of Terrestrial Magnetism, Carnegie Institution for Science, 5241 Broad Branch Road, NW, Washington, DC 20015, USA}
\newcommand{\UPenn}{The University of Pennsylvania, Department of Physics and Astronomy, Philadelphia, PA, 19104, USA}
\newcommand{\montana}{Department of Physics and Astronomy, University of Montana, 32 Campus Drive, No. 1080, Missoula, MT 59812 USA}
\newcommand{\psu}{Department of Astronomy \& Astrophysics, The Pennsylvania State University, 525 Davey Lab, University Park, PA 16802, USA}
\newcommand{\psust}{Center for Exoplanets and Habitable Worlds, The Pennsylvania State University, 525 Davey Lab, University Park, PA 16802, USA}
\newcommand{\Kutztown}{Department of Physical Sciences, Kutztown University, Kutztown, PA 19530, USA}
\newcommand{\udel}{Department of Physics \& Astronomy, University of Delaware, Newark, DE 19716, USA}
\newcommand{\Westminster}{Department of Physics, Westminster College, New Wilmington, PA 16172}
\newcommand{\steward}{Department of Astronomy and Steward Observatory, University of Arizona, Tucson, AZ 85721, USA}
\newcommand{\saao}{South African Astronomical Observatory, PO Box 9, Observatory, 7935, Cape Town, South Africa}
\newcommand{\salt}{Southern African Large Telescope, PO Box 9, Observatory, 7935, Cape Town, South Africa}
\newcommand{\ssl}{Societ\`{a} Astronomica Lunae, Italy}
\newcommand{\spot}{Spot Observatory, Nashville, TN 37206, USA}
\newcommand{\txamGP}{George P.\ and Cynthia Woods Mitchell Institute for Fundamental Physics and Astronomy, Texas A\&M University, College Station, TX77843 USA}
\newcommand{\txam}{Department of Physics and Astronomy, Texas A\&M university, College Station, TX 77843 USA}
\newcommand{\wellesley}{Department of Astronomy, Wellesley College, Wellesley, MA 02481, USA}
\newcommand{\byu}{Department of Physics and Astronomy, Brigham Young University, Provo, UT 84602, USA}
\newcommand{\Hazelwood}{Hazelwood Observatory, Churchill, Victoria, Australia}
\newcommand{\pest}{Perth Exoplanet Survey Telescope}
\newcommand{\Winer}{Winer Observatory, PO Box 797, Sonoita, AZ 85637, USA}
\newcommand{\icpo}{Ivan Curtis Private Observatory}
\newcommand{\elsauce}{El Sauce Observatory, Chile}
\newcommand{\crow}{Atalaia Group \& CROW Observatory, Portalegre, Portugal}
\newcommand{\dfus}{Dipartimento di Fisica ``E.R.Caianiello'', Universit\`a di Salerno, Via Giovanni Paolo II 132, Fisciano 84084, Italy}
\newcommand{\indfn}{Istituto Nazionale di Fisica Nucleare, Napoli, Italy}
\newcommand{\sotes}{Gabriel Murawski Private Observatory (SOTES)}
\newcommand{\lco}{Las Cumbres Observatory Global Telescope, 6740 Cortona Dr., Suite 102, Goleta, CA 93111, USA}
\newcommand{\ucsb}{Department of Physics, University of California, Santa Barbara, CA 93106-9530, USA}
\newcommand{\yale}{Department of Astronomy, Yale University, 52 Hillhouse Avenue, New Haven, CT 06511, USA}
\newcommand{\eso}{European Southern Observatory, Alonso de C\'ordova 3107, Vitacura, Casilla 19001, Santiago, Chile}
\newcommand{\stsci}{Space Telescope Science Institute, Baltimore, MD 21218, USA}
\newcommand{\keele}{Astrophysics Group, Keele University, Staffordshire ST5 5BG, UK}
\newcommand{\gsfcsellers}{GSFC Sellers Exoplanet Environments Collaboration, NASA Goddard Space Flight Center, Greenbelt, MD 20771 }
\newcommand{\usno}{U.S. Naval Observatory, Washington, DC 20392, USA}
\newcommand{\kansas}{Department of Physics and Astronomy, University of Kansas, 1251 Wescoe Hall Dr., Lawrence, KS 66045, USA}
\newcommand{\gmu}{George Mason University, 4400 University Drive MS 3F3, Fairfax, VA 22030, USA}
\newcommand{\unsw}{Exoplanetary Science at UNSW, School of Physics, UNSW Sydney, NSW 2052, Australia}
\newcommand{\sifa}{School of Physics, Sydney Institute for Astronomy (SIfA), The University of Sydney, NSW 2006, Australia}
\newcommand{\nanjing}{School of Astronomy and Space Science, Key Laboratory of Modern Astronomy and Astrophysics in Ministry of Education, Nanjing University, Nanjing 210046, Jiangsu, China}
\newcommand{\jhuapl}{Johns Hopkins APL, 11100 Johns Hopkins Rd, Laurel, MD 20723, USA}
\newcommand{\torres}{\altaffiliation{Juan Carlos Torres Fellow}}
\newcommand{\sagan}{\altaffiliation{NASA Sagan Fellow}}
\newcommand{\bernoulli}{\altaffiliation{Bernoulli fellow}}
\newcommand{\gruber}{\altaffiliation{Gruber fellow}}
\newcommand{\kavli}{\altaffiliation{Kavli Fellow}}
\newcommand{\peg}{\altaffiliation{51 Pegasi b Fellow}}
\newcommand{\pappalardo}{\altaffiliation{Pappalardo Fellow}}
\newcommand{\hubble}{\altaffiliation{NASA Hubble Fellow}}
\newcommand{\nsf}{\altaffiliation{National Science Foundation Graduate Research Fellow}}

\correspondingauthor{Noah Vowell} 
\email{vowellno@msu.edu}

\author[0000-0002-0701-4005]{Noah Vowell} 
\affiliation{\msu}

\author[0000-0001-8812-0565]{Joseph E. Rodriguez} 
\affiliation{\msu}

\author[0000-0002-8964-8377]{Samuel N. Quinn} 
\affiliation{\cfa}

\author[0000-0002-4891-3517]{George Zhou} 
\affiliation{\usq}

\author[0000-0001-7246-5438]{Andrew Vanderburg} 
\affiliation{\MIT}


\author[0000-0003-3654-1602]{Andrew W. Mann}%
\affiliation{Department of Physics and Astronomy, The University of North Carolina at Chapel Hill, Chapel Hill, NC 27599, USA}

\author[0000-0003-0030-332X]{Matthew J. Hooton}
\affiliation{Cavendish Laboratory, JJ Thomson Avenue, Cambridge, CB3 0HE, UK}

\author[0000-0002-3481-9052]{Keivan G. Stassun} 
\affiliation{\vanderbilt}
\affiliation{\fisk}

\author[0000-0003-4894-7271]{Saburo Howard}
\affiliation{Universit\'e C\^ote d'Azur, Observatoire de la C\^ote d'Azur, CNRS, Lagrange Laboratory, Nice, France}

\author[0000-0001-6637-5401]{Allyson Bieryla} 
\affiliation{\cfa}

\author[0000-0001-9911-7388]{David W. Latham} 
\affiliation{\cfa}

\author[0000-0002-2532-2853]{Steve~B.~Howell}
\affil{\ames}

\author[0000-0002-7188-8428]{Tristan Guillot}
\affiliation{Universit\'e C\^ote d'Azur, Observatoire de la C\^ote d'Azur, CNRS, Lagrange Laboratory, Nice, France}

\author{Carl Ziegler}
\affiliation{Department of Physics, Engineering and Astronomy, Stephen F. Austin State University, 1936 North St, Nacogdoches, TX 75962, USA}

\author[0000-0001-6588-9574]{Karen A.\ Collins}
\affiliation{\cfa}

\author[0000-0001-6416-1274]{Theron W. Carmichael}
\affiliation{Institute for Astronomy, University of Edinburgh, Royal Observatory, Blackford Hill, Edinburgh, EH9 3HJ, UK}

\author[0000-0002-4715-9460]{Jon M. Jenkins}
\affiliation{\ames}

\author[0000-0002-1836-3120]{Avi Shporer}
\affiliation{\MIT}

\author[0000-0002-0856-4527]{Lyu ABE}
\affiliation{Universit\'e C\^ote d'Azur, Observatoire de la C\^ote d'Azur, CNRS, Lagrange Laboratory, Nice, France}

\author{Philippe Bendjoya}
\affiliation{Universit\'e C\^ote d'Azur, Observatoire de la C\^ote d'Azur, CNRS, Lagrange Laboratory, Nice, France}


\author[0000-0002-9446-9250]{Jonathan L. Bush}%
\affiliation{Department of Physics and Astronomy, The University of North Carolina at Chapel Hill, Chapel Hill, NC 27599, USA} 

\author[0000-0003-4559-049X]{Marco Buttu}
\affiliation{INAF - Osservatorio Astronomico di Cagliari, Cagliari, Italy}

\author[0000-0003-2781-3207]{Kevin I.\ Collins}
\affiliation{George Mason University, 4400 University Drive, Fairfax, VA, 22030 USA}

\author[0000-0003-3773-5142]{Jason D. Eastman}
\affiliation{\cfa}

\author[0000-0002-9641-3138]{Matthew J. Fields}
\affiliation{Department of Physics and Astronomy, The University of North Carolina at Chapel Hill, Chapel Hill, NC 27599, USA}

\author[0000-0002-7913-4866]{Thomas Gasparetto}
\affiliation{Institute of Polar Sciences - CNR, via Torino, 155 - 30172 Venice-Mestre, Italy}


\author[0000-0002-3164-9086]{Maximilian N. G{\"u}nther}
\affiliation{European Space Agency (ESA), European Space Research and Technology Centre (ESTEC), Keplerlaan 1, 2201 AZ Noordwijk, Netherlands}

\author{Veselin B. Kostov}
\affiliation{NASA Goddard Space Flight Center, 8800  Greenbelt Road, Greenbelt, MD 20771, USA}
\affiliation{SETI Institute, 189 Bernardo Ave, Suite 200, Mountain view, CA 94043, USA}

\author[0000-0001-9811-568X]{Adam L. Kraus}%
\affiliation{Department of Astronomy, The University of Texas at Austin, Austin, TX 78712, USA}


\author[0000-0002-9903-9911]{Kathryn V. Lester}
\affil{\ames}

\author[0000-0001-8172-0453]{Alan M. Levine}
\affiliation{\MIT}

\author[0000-0001-7746-5795]{Colin Littlefield}
\affiliation{Bay Area Environmental Research Institute, Moffett Field, CA 94035, USA}
\affiliation{\ames}

\author{Wenceslas Marie-Sainte}
\affiliation{Institut Paul Emile Victor, Concordia Station, Antarctica}

\author{Djamel M\'{e}karnia}
\affiliation{Universit\'e C\^ote d'Azur, Observatoire de la C\^ote d'Azur, CNRS, Lagrange Laboratory, Nice, France}

\author{Hugh P. Osborn}
\affiliation{Physikalisches Institut, University of Bern, Gesellsschaftstrasse 6, 3012 Bern, Switzerland}
\affiliation{\MIT}


\author[0000-0003-2196-6675]{David Rapetti}
\affiliation{\ames}
\affiliation{Research Institute for Advanced Computer Science, Universities Space Research Association, Washington, DC 20024, USA}

\author[0000-0003-2058-6662]{George R. Ricker}
\affiliation{\MIT}

\author[0000-0002-6892-6948]{S. Seager}
\affil{Department of Earth, Atmospheric, and Planetary Sciences, Massachusetts Institute of Technology, Cambridge, MA 02139, USA}
\affil{\MIT}
\affil{Department of Aeronautics and Astronautics, Massachusetts Institute of Technology, Cambridge, MA 02139, USA}

\author{Gregor Srdoc}
\affiliation{Kotizarovci Observatory, Sarsoni 90, 51216 Viskovo, Croatia}

\author{Olga Suarez}
\affiliation{Universit\'e C\^ote d'Azur, Observatoire de la C\^ote d'Azur, CNRS, Lagrange Laboratory, Nice, France}

\author[0000-0002-5286-0251]{Guillermo Torres}
\affiliation{\cfa}

\author[0000-0002-5510-8751]{Amaury H.M.J. Triaud}
\affiliation{School of Physics \& Astronomy, University of Birmingham, Edgbaston, Birmingham B15 2TT, United Kingdom}

\author[0000-0001-6763-6562]{R.~Vanderspek}
\affiliation{\MIT}

\author[0000-0002-4265-047X]{Joshua N.\ Winn}
\affiliation{Department of Astrophysical Sciences, Princeton University, Princeton, NJ 08544, USA}

\shorttitle{HIP 33609 b}
\shortauthors{Vowell et al.}

\begin{abstract}
We present the discovery and characterization of \thisplanet, a transiting warm brown dwarf orbiting a late B-star, discovered by NASA’s Transiting Exoplanet Survey Satellite (\tess) as TOI-588 b. \thisplanet is a large (\rplanet) brown dwarf on a highly eccentric (\ecc) orbit with a 39-day period. The host star is a bright (V = 7.3 mag), \teffstar\ star with a mass of \starmass\ and radius of \starrad, making it the hottest transiting brown dwarf host star discovered to date. We obtained radial velocity measurements from the CHIRON spectrograph confirming the companion’s mass of \mplanet\ as well as the host star’s rotation rate ($\vsini = 55.6 \pm 1.8$ km/s). We also present the discovery of a new comoving group of stars, designated as MELANGE-6, and determine that   \thisstar is a member. We use a combination of rotation periods and isochrone models fit to the cluster members to estimate an age of 150 $\pm$ 25 Myr. With a measured mass, radius, and age, \thisplanet becomes a benchmark for substellar evolutionary models.

\end{abstract}

\section{Introduction}
\label{sec:introduction}
Brown dwarfs (BDs), defined as objects that fuse only deuterium at some point in their lifetime, occupy the region in mass between planets and stars. The mass range corresponding to this historical definition runs between a lower limit of 11-16 Jupiter Masses (MJ), where an object begins to fuse deuterium \citep{Spiegel2011} and an upper limit of ~75-80 MJ where hydrogen fusion begins \citep{Baraffe2002}. However, these fusion based transitions that distinguish BDs from planets and stars may occur at masses that depend on multiple factors. Specifically, the environment that the object formed in, the effects of convection on the object, and its metallicity can influence these traditional mass limits \citep{Spiegel2011}. Therefore, it may be preferable to define BDs in the context of their formation and evolution, an idea suggested by \citet{Chabrier2014}, \citet{Burrows2001}, and \citet{Carmichael2021}. 

It is likely that some BDs form and evolve in similar ways to giant planets, providing the opportunity for comparative studies with the known exoplanet sample and possibly gaining insight into the evolutionary pathways of BDs. We know that planets can migrate through quiet mechanisms like disk-driven migration \citep{DAngelo:2003} as well as dynamical interactions such as planet-planet scattering or Kozai-Lidov cycles \citep{Kozai1962, Lidov1962, Fabrycky:2007}. Such dynamical interactions lead to a fraction of the population residing in close-in, highly misaligned as well as highly eccentric orbits \citep{Rasio:1996, Wu:2011}. We can measure these misalignments using Doppler tomography \citep{CollierCameron:2010,Zhou:2016} and the Rossiter-McLaughlin effect \citep{Rossiter:1924, McLaughlin:1924, Queloz2010}. These techniques constrain the orbital obliquity of the companion by tracking the effects of the transiting planet’s shadow on the rotationally broadened stellar spectral line profile or on the apparent radial velocity of the host star. If BDs are indeed subject to the same dynamical interactions as planets, then we should expect to see similar signatures in the transiting BD population.

In order to fully understand the formation and evolutionary history of BDs, we need robust, well-tested models for substellar evolution. Current substellar evolutionary models show that BDs rapidly contract in the first billion years after formation \citep{Baraffe2003, SaumonMarley2008, Burrows2001, Phillips2020}, and then continue slowly contracting out to $\sim$10 Gyr. Combining this with the fact that BDs are held up by electron degeneracy pressure and therefore, at late times, tend to have radii that decrease with mass reveals the significance of obtaining precise age estimates for transiting BDs. Thus, in order to test our models of substellar evolution, we need precise, independent measurements of BD radii, masses, and ages. Through the combined efforts of NASA's Transiting Exoplanet Survey Satellite (\tess; \citealp{Ricker:2015}), ground-based follow-up programs, and ESA's Gaia mission \citep{Gaia2022} we are able to obtain precise measurements of transiting BD radii and masses with transit photometry, radial velocity measurements, spectral energy distributions (SEDs), and Gaia parallaxes.

Precisely measuring the age of the BD, while difficult, is vital to furthering our understanding of BD evolution because it provides a direct test of substellar evolutionary models. Unfortunately, only 4 of the 37 published transiting BD systems have precisely measured ages ($>3\sigma$) determined either through stellar cluster memberships \citep{Gillen2017,Beatty2018, David2019} or leveraging gyrochronology and lithium abundances \citep{Carmichael2021}. The simplest method for obtaining precise ages of BDs would be to discover more around host stars that are members of clusters. This kind of targeted discovery is already underway for planets by the \tess\ Hunt for Young and Maturing Exoplanets (THYME) consortium \citep{Newton2019} and can also be applied to BDs in order to better understand their evolutionary pathways.

In this paper, we present the discovery of \thisplanet from NASA's \tess\ mission. \thisplanet adds to the growing number of transiting BDs discovered by \tess\ that is approaching a population large enough to begin performing robust demographic analyses. It is also a benchmark system for testing BD formation and evolution since it has an age measurement from membership in a stellar association, and its host star is both the brightest (\textit{V} = 7.3 mag) and hottest (\teffstar) star with a transiting BD companion discovered so far. \thisplanet's high orbital eccentricity (\ecc) could be indicative of a dynamically active past, and we should therefore search for additional evidence of past interactions (such as a large stellar obliquity). In \S\ref{sec:observations} we present our follow up photometric and spectroscopic observations obtained through the \tess\ Follow-up Observing Program (TFOP) Working Group. We establish \thisstar's cluster membership and age in \S\ref{sec:ClusterMembership}. In \S\ref{sec:Analysis} we describe our global modeling methodology using \texttt{EXOFASTv2} \citep{Eastman:2013, Eastman:2019} as well a separate analysis on the effects of gravity darkening. We place \thisplanet in context with other transiting substellar companions and discuss future characterization prospects in \S\ref{sec:results}. We present our conclusions in \S\ref{sec:conclusion}.

\begin{figure}
\centering
\includegraphics[trim = 0 0 0 0,width=\linewidth]{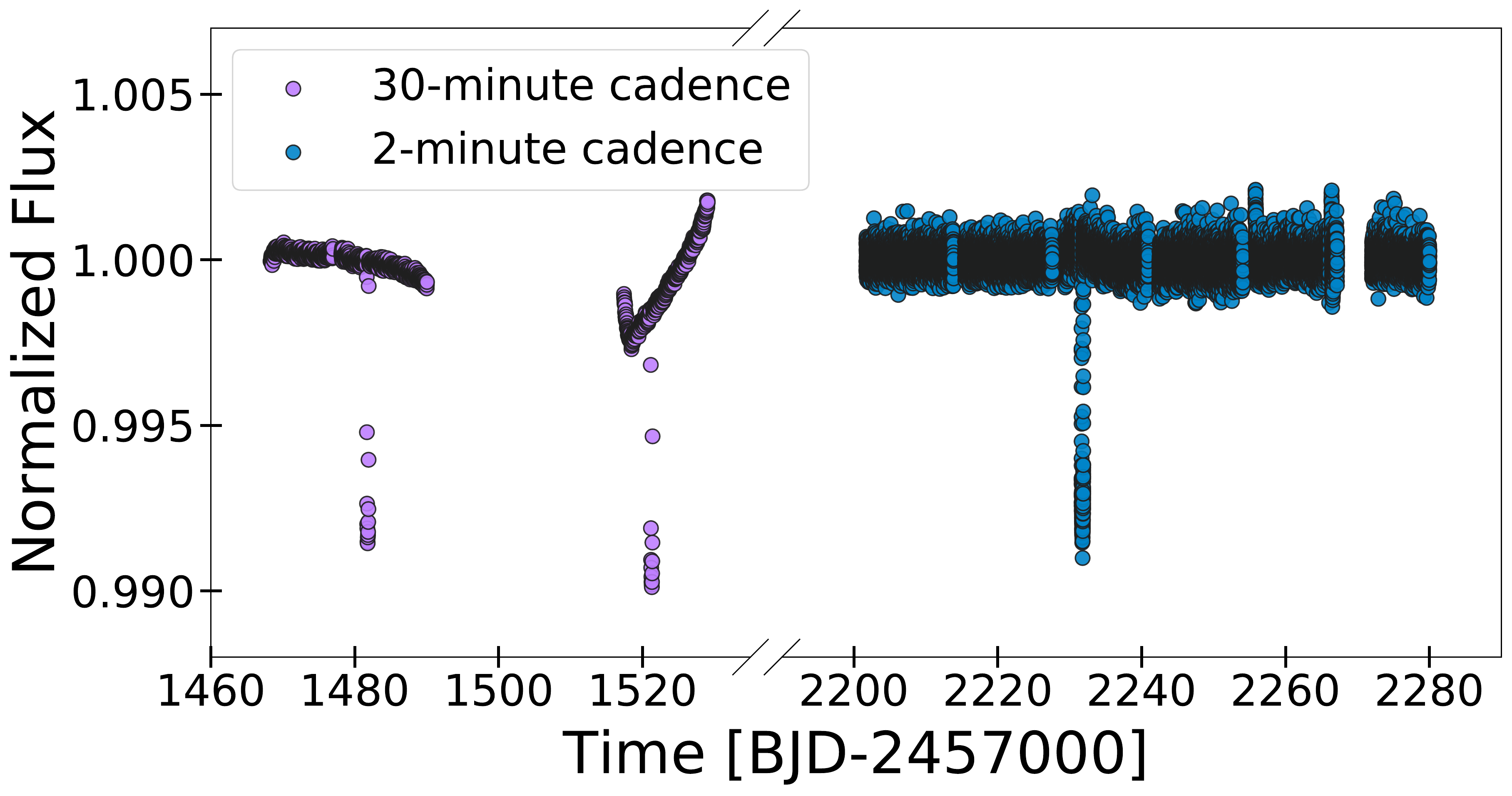}
\caption{The \tess\ light curves from sectors 6, 8, 33, 34, and 35 extracted using the techniques described in \S\ref{sec:TESS}. We note that the gap in flux located at BJD 2459270 is caused by a period of coarse pointing. These poor data were removed in our fit as described in \S\ref{sec:TESS}.}
\label{fig:fullLC}
\end{figure}

\begin{table}
\scriptsize
\setlength{\tabcolsep}{2pt}
\centering
\caption{Literature and Measured Properties for \thisstar}
\begin{tabular}{llcc}
  \hline
  \hline
Other identifiers\dotfill & \\
& TOI-588\\
& TIC 130415266\\
& HD 52470\\
& HIP 33609\\
& TYC 8122-01924-1\\
& 2MASS J06585996-4701240\\ 
&TESS Sector & [6, 8, 33, 34, 35$^*$]\\
\hline
\hline
Parameter & Description & Value & Reference\\
\hline 
$\alpha_{J2000}\ddagger$\dotfill&Right Ascension (RA)\dotfill &06:58:59.966& 1 \\
$\delta_{J2000}\ddagger$\dotfill&Declination (Dec)\dotfill &-47:01:24.121& 1 \\
${\rm G}$\dotfill     & Gaia $G$ mag.\dotfill     & 7.26$\pm$0.02& 1 \\
B$_{\rm P}$\dotfill			&Gaia B$_{\rm P}$ mag.\dotfill & 7.27$\pm$0.02& 1 \\
R$_{\rm P}$\dotfill			&Gaia R$_{\rm P}$ mag.\dotfill & 7.30$\pm$0.02& 1 \\
${\rm T}$\dotfill     & TESS mag.\dotfill     & 7.312$\pm$0.006& 2 \\
&                \\
B$_T$\dotfill			&Tycho B$_T$ mag.\dotfill & 7.271$\pm$0.02		& 3	\\
V$_T$\dotfill			&Tycho V$_T$ mag.\dotfill & 7.284$\pm$0.02		& 3	\\
&                \\
J\dotfill			& 2MASS J mag.\dotfill &7.245$\pm$0.020& 4 \\
H\dotfill			& 2MASS H mag.\dotfill & 7.326$\pm$0.031& 4 \\
K$_S$\dotfill			& 2MASS ${\rm K_S}$ mag.\dotfill &7.278$\pm$0.027& 4 \\
&                \\
\textit{WISE1}\dotfill		& \textit{WISE1} mag.\dotfill &7.263$\pm$0.036& 5 \\
\textit{WISE2}\dotfill		& \textit{WISE2} mag.\dotfill &7.326$\pm$0.030& 5 \\
\textit{WISE3}\dotfill		& \textit{WISE3} mag.\dotfill &7.354$\pm$0.030& 5 \\
\textit{WISE4}\dotfill		& \textit{WISE4} mag.\dotfill & 7.4$\pm$0.1 & 5 \\
&                \\
$\mu_{\alpha}$\dotfill		& Gaia DR3 proper motion\dotfill & -9.505$\pm$0.073& 1 \\	
                    & \hspace{3pt} in RA (mas yr$^{-1}$)	&&                \\
$\mu_{\delta}$\dotfill		& Gaia DR3 proper motion\dotfill 	& -4.467$\pm$0.071& 1 \\
                    & \hspace{3pt} in DEC (mas yr$^{-1}$) &  &                \\
&                \\
$v\sin{i_\star}$\dotfill &  Rotational velocity (\kms) \hspace{9pt}\dotfill & 55.6$\pm$1.8& \S\ref{sec:chiron}               \\
$\pi^\dagger$\dotfill & Gaia DR3 Parallax (mas) \dotfill &6.49$\pm$0.05&1\\
\hline\\
\hline\\
\end{tabular}
\begin{flushleft}
 \footnotesize{ \textbf{\textsc{NOTES:}}
 The uncertainties of the photometry have a systematic error floor applied. \\
 $\ddagger$ RA and Dec are in epoch J2000. The coordinates come from Vizier where the Gaia RA and Dec have been precessed and corrected to J2000 from epoch J2015.5.\\
 $\dagger$ Values have been corrected for the -0.30 $\mu$as offset as reported by \citet{Lindegren:2018} but this is not significant for these systems.\\
 References are: $^1$\citep{Gaia2022},$^2$\citep{Hog:2000}, $^3$\citet{Stassun:2018_TIC},$^4$\citet{Cutri:2003}, $^5$\citet{Cutri:2012}\\
}
\end{flushleft}
\label{tbl:LitProps}
\end{table}

\section{Observations}
\label{sec:observations}
\begin{figure}
\centering 
\includegraphics[trim = 0 0 0 0,width=0.99\linewidth]{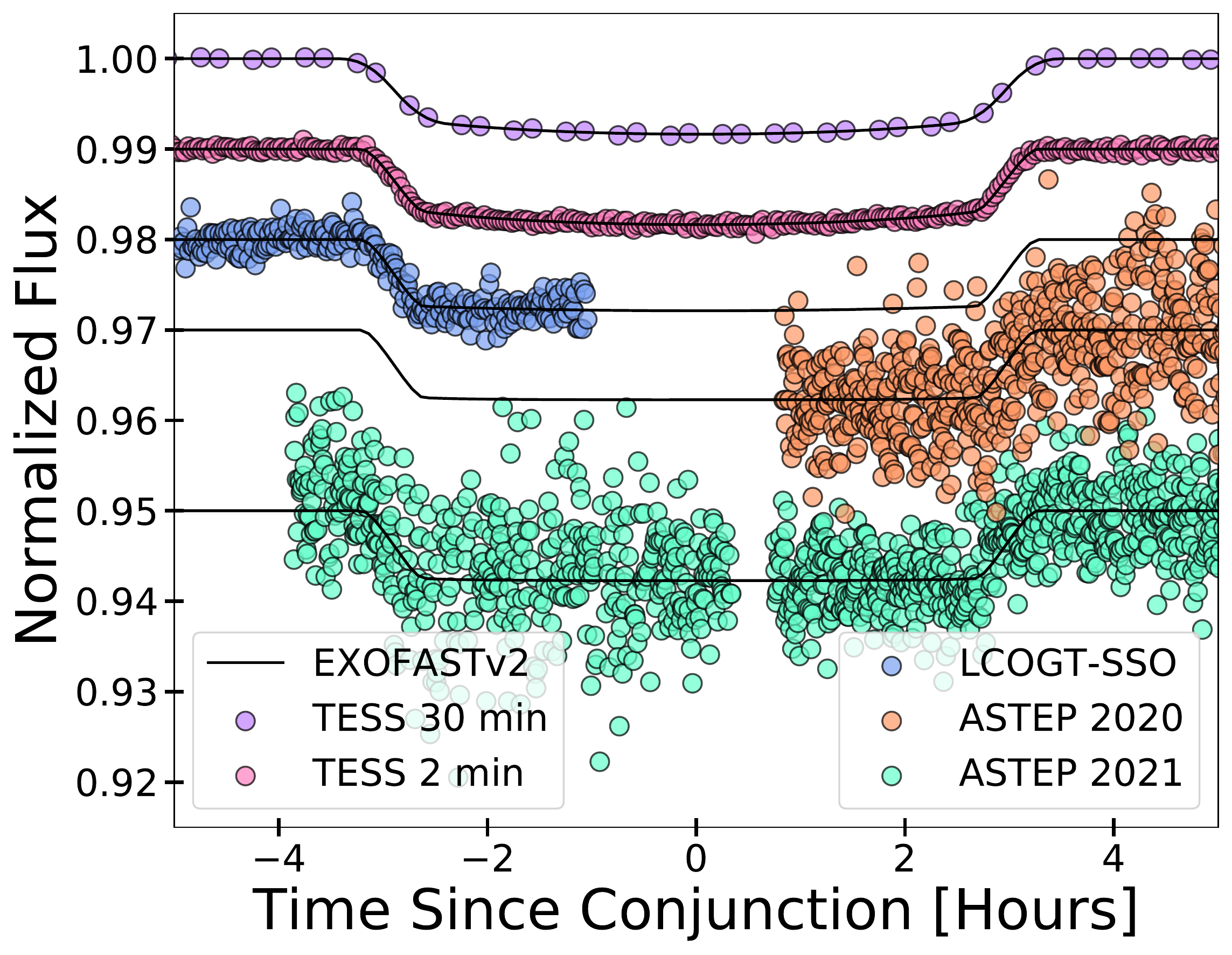}
\caption{The \tess\ and ground-based follow-up transits for \thisplanet described in sections \S\ref{sec:TESS} and \S\ref{sec:follow-up}. The model for each transit is shown as a black solid line.}
\label{fig:stack_plot}
\end{figure}

\begin{figure*}
\centering 
\includegraphics[trim = 0 0 0 0,width=\linewidth]{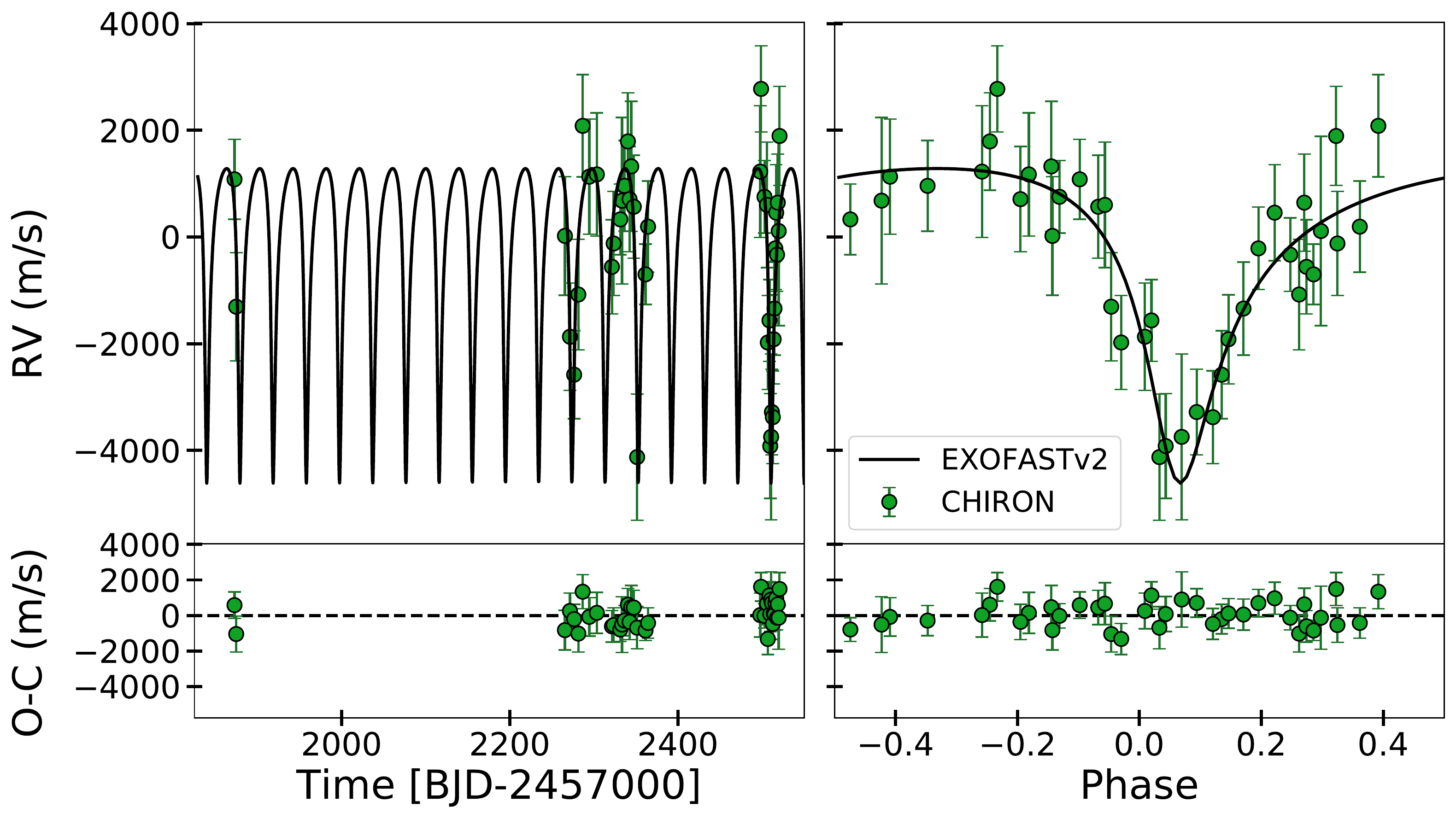}
\caption{The radial velocity observations from CHIRON unphased (left) and phased to our best fit ephemeris (right).}
\label{fig:rv_plot}
\end{figure*}

To measure the mass and orbital parameters of the \thisstar system, we used a combination of photometric (Figures \ref{fig:fullLC} and \ref{fig:stack_plot}) and spectroscopic (Figure \ref{fig:rv_plot}) observations. The observations, gathered through the \tess\ Follow-up Observing Program (TFOP), were part of the vetting process to rule out false positive scenarios. We describe these observations in the following sub-sections.

\subsection{{\it TESS} Photometry}
\label{sec:TESS}

\tess\ observes a $24^{\circ}$x$96^{\circ}$ patch of the sky for approximately 27 days before moving to a new sector \citep{Ricker:2015}. In the prime mission, it observed its entire field of view at 30-minute cadence, and a pre-selected set of stars were observed at 2-minute cadence, resulting in $>$80\% of the entire sky being observed. \tess\ just completed its first extended mission in which it observed a portion of the ecliptic plane, the region of sky observed by the \textit{K2} mission \citep{Howell2014} which repurposed the Kepler spacecraft to observe the ecliptic plane after the loss of the spacecraft's 2nd of four reaction wheels. This region was not observed during the TESS prime mission, but a portion was observed over a 5-month period in the first extended mission, and another portion will be observed over a 4-month period in the second extended mission which started 2 September 2022. A subset of $\sim$2,000 of the 20,000 preselected targets in the extended missions\footnote{\url{https://heasarc.gsfc.nasa.gov/docs/tess/the-tess-extended-mission.html}} are observed at 20-second cadence in addition to 2-minute cadence, and the exposure time for the Full Frame Images (FFI) was reduced to 10 minutes in the first extended mission, and further reduced to just 200 seconds in the second extended mission. \thisplanet was first observed during the primary mission in the sector 6 FFIs at 30-minute cadence in 2018 and then again in sector 8 in 2019. \tess\ then observed \thisstar again during its first extended mission at 2-minute cadence in 2021 during sectors 33, 34, and 35. 

\tess\ observations are downloaded, reduced, and analyzed on the ground. The original detection of a transiting signal around \thisstar\ was made by the MIT Quick Look Pipeline (QLP), and it was then vetted as a \tess\ Object of Interest (TOI-588, Table \ref{tbl:LitProps}) using the process described by \citet{Guerrero:2021}. In subsequent observations, the data collected by TESS at 2-minute cadence were processed by the Science Processing Operations Center (SPOC) pipeline \citep{Jenkins:2016} based at NASA Ames Research Center where the image data were calibrated, and light curves were extracted for each target, which were then searched for transiting planet signatures. We then downloaded these SPOC PDC-SAP light curves \citep{Smith:2012, Stumpe2012, Stumpe:2014} from the Mikulski Archive for Space Telescopes (MAST) using the \textit{Lightkurve 2} software \citep{Lightkurve2018}. The SPOC transit search over sectors 34 and 35 triggered on the single transit of HIP 33609 b in sector 34, but at the wrong period. Nevertheless, the difference image centroiding test located the source of the transit signature within 1.2 $\pm$ 2.8 arcsec. We conducted a Lomb-Scargle period search on the 2-minute, normalized light curve with the transits masked out to search for stellar rotation from star spots. We searched for periods ranging from 0.1 to 10 days and found no significant signal as expected from a relatively quiet B-type star.

We found a total of 4 transits of \thisstar in sectors 6, 8, 34, and 35. However, we discarded the sector 35 transit due to poor data quality caused by a period of unstable pointing of the \tess\ spacecraft. During this period, the stars moved around significantly on the detector, introducing large systematic errors. 

For our global analysis, we used the SPOC 2-minute light curve for the sector 34 transit, but we re-extracted the 30-minute FFI light curves for sectors 6 and 8 using a custom FFI pipeline based on the procedure described by \citet{Vanderburg2019}. In brief, we performed aperture photometry on a series of 20 apertures, decorrelated each extracted light curve against the background flux outside the aperture and the mean value and standard deviation of pointing excursions during each exposure (measured from the spacecraft quaternion time series), and selected the light curve from the aperture that maximized photometric precision. We then removed any long-term instrumental and stellar variability signals by fitting a spline to the flux using {\it Keplerspline}\footnote{\url{https://github.com/avanderburg/keplerspline}} and then dividing the light curve by the best fit model \citep{Vanderburg:2014}. We then removed most of the out-of-transit baseline from both light curves keeping only half a transit duration on each side of the transit since these data provide little to no information to the global fit while simultaneously being computationally expensive to model. These light curves were fit simultaneously with all available data on the \thisstar system (see \S\ref{sec:GlobalModel}).

\subsection{Ground-based Photometric Follow-up}
\label{sec:follow-up}
In order to rule out contamination by a background eclipsing binary and refine the ephemeris, we observed \thisstar as a part of SubGroup 1 (seeing limited photometry) of the \tess\ Follow-up Observing Program (TFOP). We obtained these observations using the Las Cumbres Observatory Global Telescope (LCOGT) telescope network \citep{Brown:2013} and the 40 cm ASTEP-400 telescope \citep{Abe2013, Guillot2015}. Both facilities confirm the existence of a transit on the target star, \thisstar, and confirm that there are no nearby stars that exhibit variability.  

We observed an ingress of \thisplanet\ on 08 December 2020 UT from LCOGT-SSO on the 1 m telescope in the y-band at 25 s cadence with a pixel scale of 0.389". We observed an egress on 12 August 2020 UT and a full transit (save for a brief gap mid-transit due to a brief noon twilight) on 23 June 2021 UT with ASTEP using a 25 s exposure time with a pixel scale of 0.93". More transit observations were attempted by LCOGT and ASTEP, however they were at too low S/N to provide value to our global fit while increasing the computational cost. We therefore discard these transits and only show the 3 ground based transits used in our analysis. We reduced these data sets and extracted the light curves using \texttt{AstroImageJ} \citep{Collins:2017}. We detrended the LCO light curve against airmass and detrended the ASTEP light curves against both airmass and sky/pixels. See \S D in the appendix of \citet{Collins:2017} for a detailed description of the detrending parameters.   

\subsection{CHIRON Spectroscopy}
\label{sec:chiron}

\begin{deluxetable}{l l l | l l l}[bt]
\tabletypesize{\scriptsize}
\tablecaption{The radial velocity measurements for the HIP 33609 system. \label{tbl:rv_data}}
\tablewidth{0pt}
\tablehead{
\colhead{\bjdtdb} & \colhead{RV (m s$^{-1}$)} & \colhead{$\sigma_{RV}$ (m s$^{-1}$)} & \colhead{\bjdtdb} & \colhead{RV (m s$^{-1}$)} & \colhead{$\sigma_{RV}$ (m s$^{-1}$)}}
\startdata
2458872.64968 & 32201.9 & 769.6 & 2459364.44268 & 31314.1 & 873.7 \\
2458874.68716 & 29813.7 & 1029.1 & 2459497.86645 & 32346.2 & 1247.6 \\
2459265.59795 & 31140.8 & 1125.7 & 2459498.86337 & 33896.8 & 823.9 \\
2459271.58653 & 29249.9 & 1019.7 & 2459502.88700 & 31873.6 & 702.7 \\
2459276.56080 & 28538.2 & 842.3 & 2459505.82903 & 31722.4 & 1189.3 \\
2459281.57067 & 30041.5 & 1049.7 & 2459506.88278 & 29142.3 & 896.2 \\
2459286.69996 & 33204.2 & 974.2 & 2459508.84044 & 29556.2 & 785.5 \\
2459294.55752 & 32251.1 & 1093.2 & 2459509.76436 & 27200.5 & 998.3 \\
2459303.54558 & 32293.7 & 1167.0 & 2459510.79825 & 27373.0 & 1564.7 \\
2459321.52137 & 30560.4 & 900.5 & 2459511.77084 & 27838.5 & 820.2 \\
2459323.52343 & 30998.6 & 992.5 & 2459512.81732 & 27742.1 & 886.0 \\
2459331.46144 & 31450.3 & 685.8 & 2459513.82850 & 29202.6 & 853.4 \\
2459333.48087 & 31801.6 & 1571.2 & 2459514.80241 & 29779.1 & 889.0 \\
2459336.46010 & 32079.4 & 868.0 & 2459515.76514 & 30906.6 & 792.1  \\
2459340.49478 & 32912.0 & 930.0  & 2459516.81742 & 31576.8 & 917.3  \\
2459342.47356 & 31829.6 & 1002.2 &2459517.82665 & 30789.3 & 711.5 \\
2459344.46994 & 32443.8 & 1230.7 & 2459518.73563 & 31764.7 & 927.9 \\
2459347.50808 & 31687.2 & 979.7 & 2459519.81360 & 31232.3 & 1781.9 \\
2459351.46803 & 26993.6 & 1195.5 & 2459520.79849 & 33014.2 & 944.1 \\
2459361.44832 & 30419.7 & 591.6 & & & \\
\enddata
\end{deluxetable}

We observed \thisstar on 39 separate nights from 24 January 2020 UT through 2 November 2021 UT (Table \ref{tbl:rv_data}) using the CHIRON spectrograph on the 1.5 m SMARTS telescope located at the Cerro Tololo Inter-American Observatory (CTIO) in Chile \citep{Tokovinin2013, Paredes2021}. CHIRON is a high resolution echelle spectrograph fed with an image slicer through a single multi-mode fiber which achieves a spectral resolving power of R $=$ 80,000 over the range 410 to 870 nm. We used these spectra to constrain the stellar parameters of the host star and extract the radial velocities (RVs). In order to extract the RVs, we derived the line profiles from our observed spectra by performing a least squares deconvolution  \citep{Donati:1997, Zhou2020}. We deconvolved against synthetic spectral templates generated using the ATLAS9 model atmospheres \citep{Kurucz:1992} with our rotational broadening kernel applied.

We measured the projected rotational velocity $\vsini$ of the host star by modeling the line profiles from our spectra with a convolution of kernels as prescribed by \citet{Zhou2018}. They consisted of rotation and radial-tangential macroturbulence kernels from \citet{Gray:2005} and an instrumental broadening kernel which is represented as a Gaussian with a width equivalent to the instrumental resolution. From this analysis, we found that \thisstar has a projected rotational velocity of $\vsini = 55.6 \pm 1.8$ km/s. We constrained the stellar atmospheric parameters such as metallicity and effective temperature by comparing our spectra to an interpolated library of spectra classified by the Stellar Parameter Classification (SPC) package \citep{Buchhave:2012}. However, we only utilized this analysis as a consistency check as determining stellar parameters through spectra is highly uncertain for rapidly rotating B-type stars like \thisstar \citep{Gaudi:2017}. Therefore, we adopt the stellar parameters derived in our global fit which are constrained by simultaneously fitting to the SED and stellar isochrones.

\subsection{Spectral Energy Distribution}
\label{sec:sed}
\begin{figure*}
    \centering
    \includegraphics[trim={0cm 0cm 0cm 0.6cm},clip,scale=0.4]{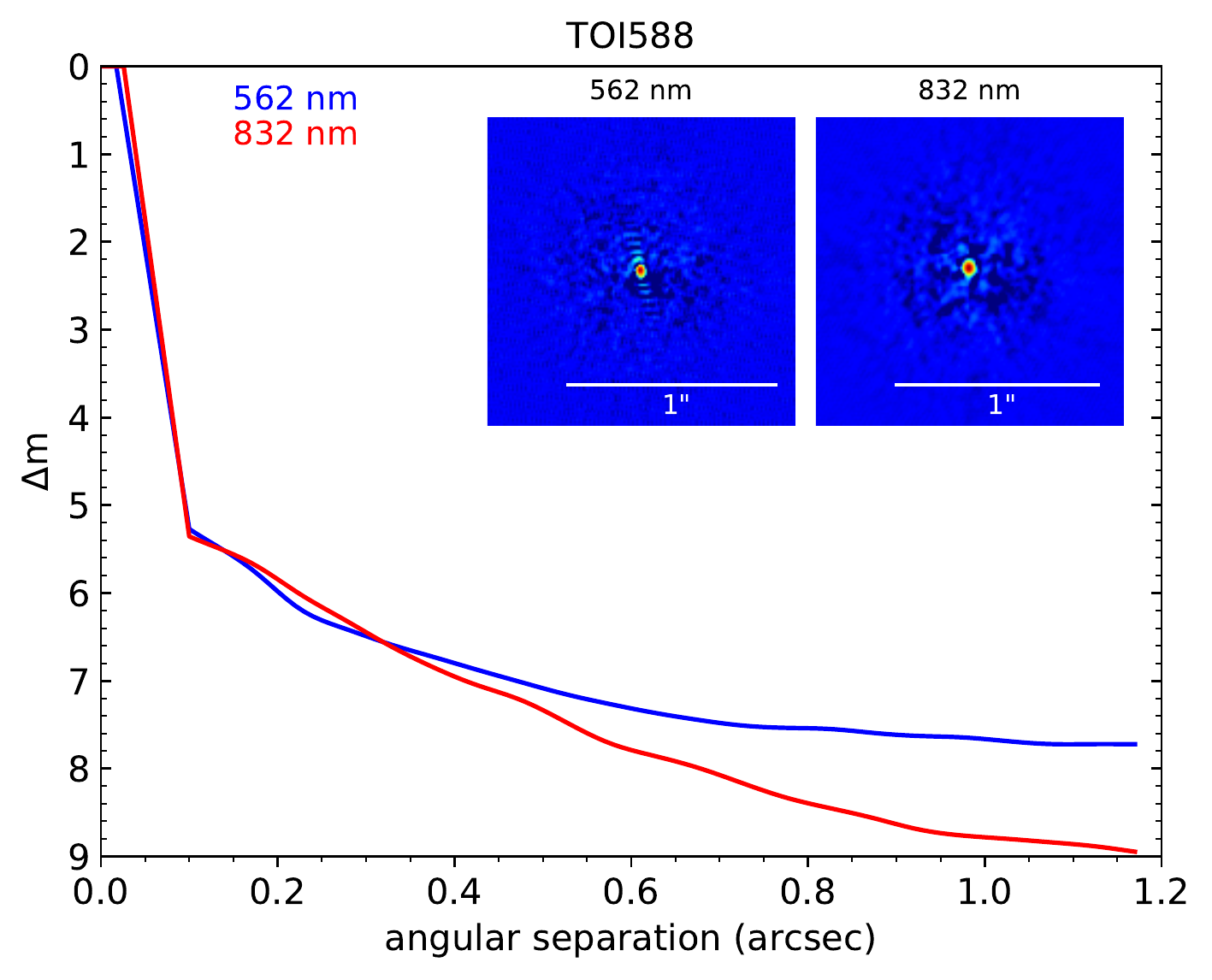}\includegraphics[trim={2cm 2.5cm 2cm 2cm},clip,scale=0.275]{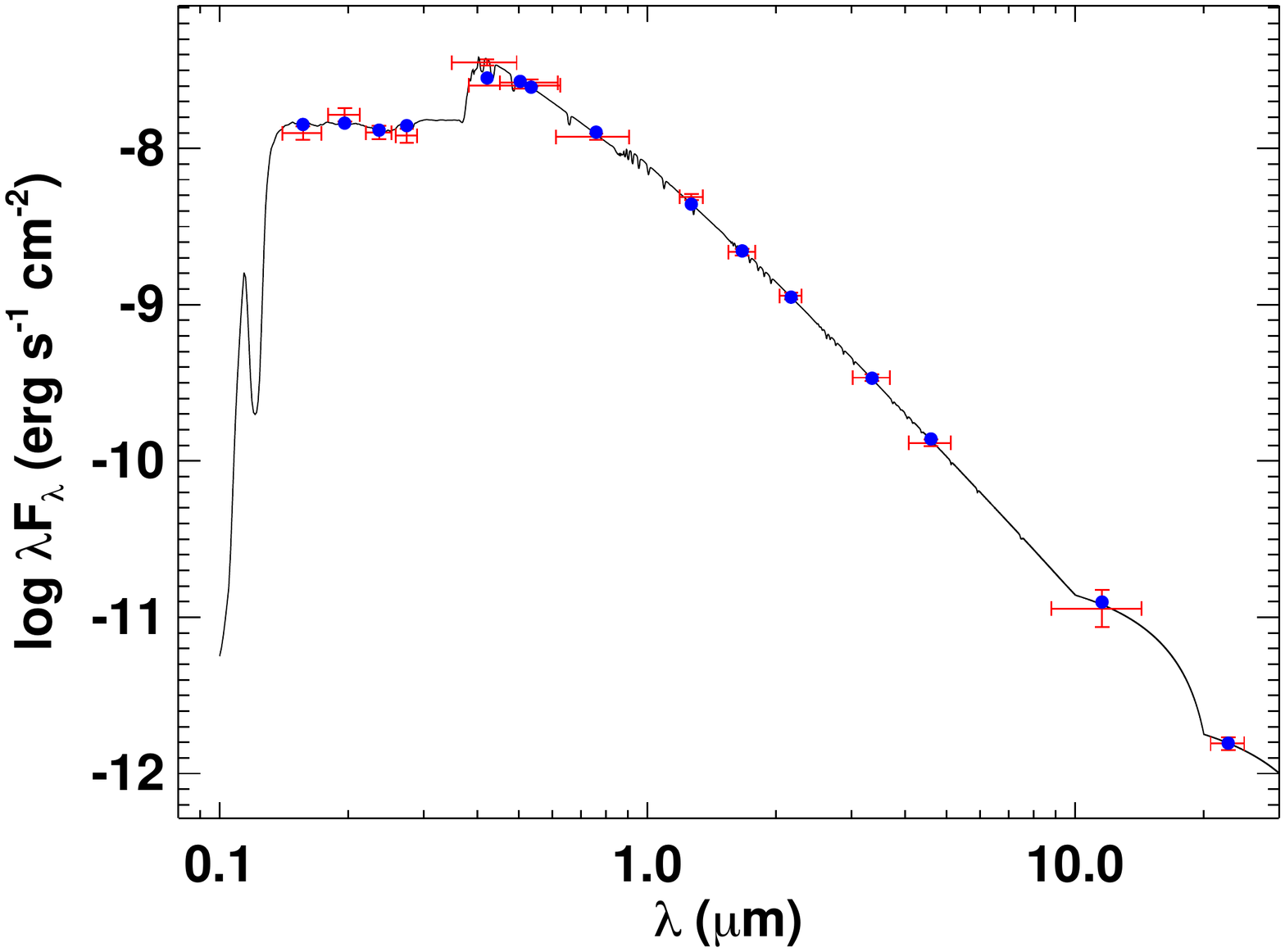}\includegraphics[trim={1.8cm 0.65cm 0cm 1.1cm},clip,scale=0.33]{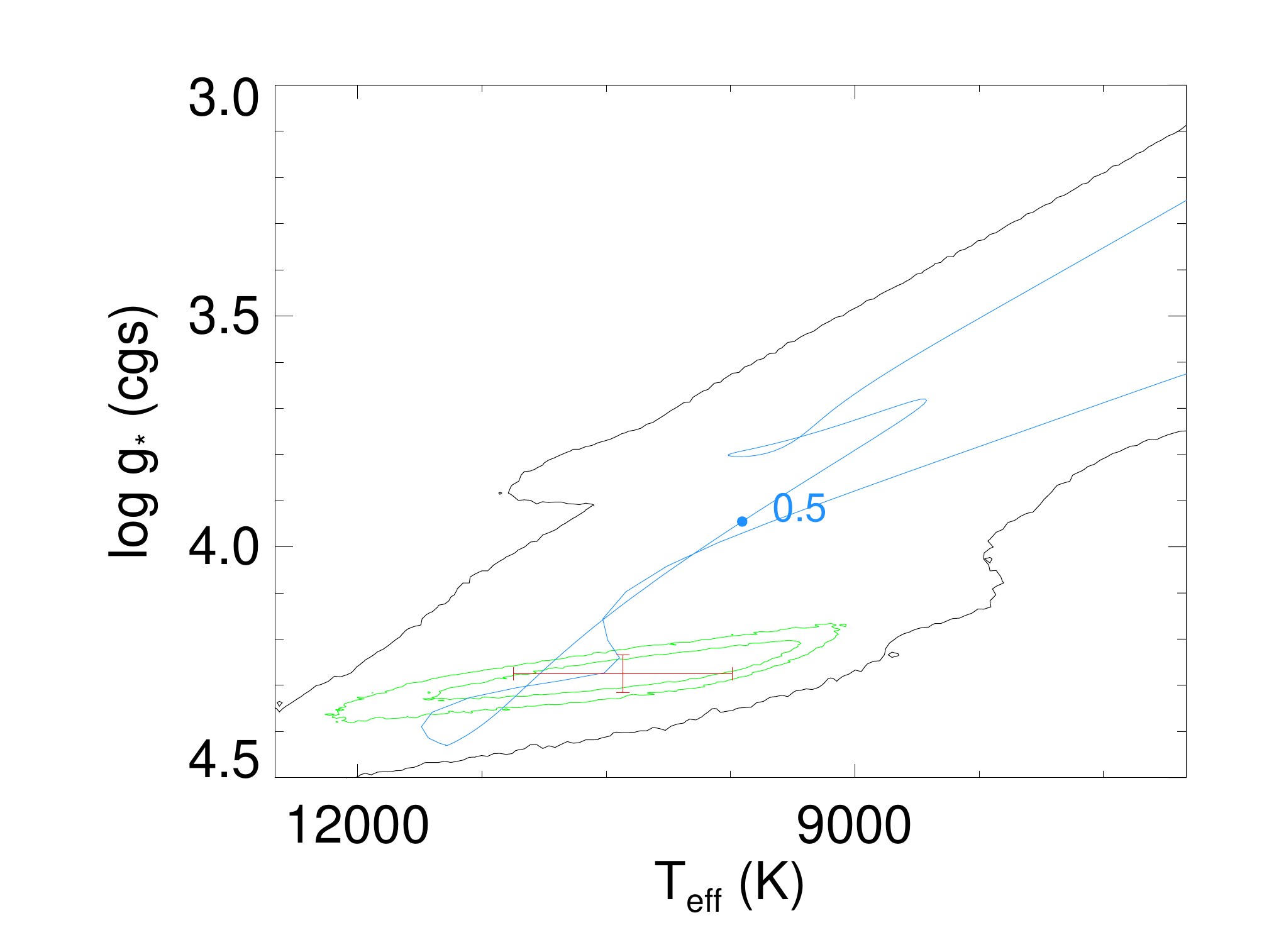}

    \caption{(Left) The 5$\sigma$ speckle imaging contrast curves in both filters as a function of the angular separation from the diffraction limit (20 mas) out to 1.2 arcsec, the end of speckle coherence. The inset shows the reconstructed 562 nm and 832 nm images with a 1 arcsec scale bar. \thisstar was found to have no close companions to within the angular and contrast levels achieved. (Middle) Spectral energy distribution of \thisstar. Red symbols represent the observed photometric measurements, where the horizontal bars represent the effective width of the passband. Blue symbols are the model fluxes from the best-fit Kurucz atmosphere model (black). (Right) The best fitting MIST evolutionary track shown in blue with the 3$\sigma$ contours on the best fit MIST track in black. The red point indicates the median value and $1\sigma$ error bars from our global analysis, while the green contours show the $3\sigma$ errors. The blue point indicates the location of 0.5 Gyr on the evolutionary track.}
    \label{fig:hires+sed}
\end{figure*}

We fit the broadband SED simultaneously as a part of our global \texttt{EXOFASTv2} analysis (see \S \ref{sec:GlobalModel}). However, we also performed a separate analysis of the SED in order to independently determine the basic stellar parameters and serve as a consistency check for our global analysis. We analyzed the SED of the star together with the {\it Gaia\/} DR3 parallax \citep[with no systematic offset applied; see, e.g.,][]{StassunTorres2021}, in order to determine an empirical measurement of the stellar radius, following the procedures described in \citet{Stassun2016,Stassun2017,Stassun:2018}. We pulled the $B_T V_T$ magnitudes from {\it Tycho-2}, the $JHK_S$ magnitudes from {\it 2MASS}, the W1--W4 magnitudes from {\it WISE}, and the $G$ $G_{\rm BP}$ $G_{\rm RP}$ magnitudes from {\it Gaia}. We also used the UV measurements at 157--274~nm from the TD1 UV satellite \citep{Boksenberg1973, Thompson1978}. Together, the available photometry spans the full stellar SED over the wavelength range 0.15--22~$\mu$m (see Figure~\ref{fig:hires+sed}).

We performed a fit to the SED using Kurucz stellar atmosphere models, with the main parameters being the effective temperature ($T_{\rm eff}$), surface gravity ($\log g$), and metallicity ([Fe/H]), for which we adopted the spectroscopically determined values: $T_{\rm eff} = 10,570^{+850}_{-710}$, $\log g = 4.259^{+0.057}_{-0.063}$, [Fe/H]$ = -0.26^{+0.38}_{-0.61}$. The remaining free parameter was the extinction $A_V$, which we limited to the maximum line-of-sight value from the Galactic dust maps of \citet{Schlegel:1998}. The resulting fit has a reduced $\chi^2$ of 1.5 and best fit $A_V = 0.20 \pm 0.03$. Integrating the model SED gives the bolometric flux at Earth, $F_{\rm bol} = 5.12 \pm 0.24 \times 10^{-8}$ erg~s$^{-1}$~cm$^{-2}$. Taking the $F_{\rm bol}$ and $T_{\rm eff}$ together with the {\it Gaia\/} parallax yields the stellar radius, $R_\star = 1.81 \pm 0.27$~R$_\odot$. In addition, we used the empirical relations of \citet{Torres:2010}, to estimate the stellar mass $M_\star = 2.38 \pm 0.14$~M$_\odot$, which is consistent with the value of $2.18 \pm 0.66$~M$_\odot$ determined empirically via $R_\star$ and $\log g$.

Finally, we can extrapolate the model atmosphere below 0.1~nm \citep[see][]{Stassun2016} to estimate the XUV radiation in the BD's environment, for which we find $F_{\rm XUV} = 302^{+532}_{-208}$ erg~s$^{-1}$~cm$^{-2}$ at a distance of 1~AU from the star. Overall, we find this analysis is consistent at the 1$\sigma$ level with the results of our global analysis, and so we adopt the results of our \texttt{EXOFASTv2} fit since it simultaneously fits all available data.

\subsection{High Resolution Imaging}
\label{sec:hires_imaging}
If an exoplanet host star has a spatially close companion, that companion (bound or line of sight) can create a false- positive transit signal if it is, for example, an eclipsing binary (EB). The flux from a close companion star constitutes "third-light” and may lead to an underestimated planetary radius if not accounted for in the transit model \citep{Ciardi:2015} and cause non-detections of small planets in the same exoplanetary system \citep{Lester2021}. Additionally, the discovery of close, bound companion stars, which exist in nearly one-half of FGK type stars \citep{Matson2018}, provides crucial information toward our understanding of exoplanetary formation, dynamics and evolution \citep{Howell2021}. Thus, to search for close-in bound companions unresolved in TESS observations, we obtained high-resolution imaging speckle observations of \thisstar.

\thisstar was observed on 2022 March 03 UT using the Zorro speckle instrument on the Gemini South 8-m telescope \citep{Scott2021, Howell2022}. Zorro provides simultaneous speckle imaging in two bands (562 nm and 832 nm) with output data products including a reconstructed image with robust contrast limits on companion detections. Three sets of 1000 $\times$ 0.06 s exposures were collected on \thisstar and subjected to Fourier analysis in our standard reduction pipeline (see \citealt{Howell2011}). Figure \ref{fig:hires+sed} shows our final contrast curves and the two reconstructed speckle images. We find that \thisstar is a single star with no companion brighter than 5-9 magnitudes below that of the target star from the diffraction limit (20 mas) out to 1.2$\arcsec$. At the distance of \thisstar (d=154 pc) these angular limits correspond to spatial limits of 3 to 185 AU.

We also observed \thisstar on 18 May 2019 UT from the 4.1-m Southern Astrophysical Research (SOAR) telescope with speckle interferometry in the \textit{I}-band \citep{Tokovinin:2018}. We took these observations in-line with the general observing strategy for \tess\ targets described in \citet{Ziegler2020} with an estimated contrast of $\Delta$mag = 7.7 at 1". We found no nearby companions out to 3".

\section{Membership and Age Determination of \thisstar}
\label{sec:ClusterMembership}

\subsection{\association}

We searched for evidence that \thisstar{} is a member of a young stellar association using the \texttt{FriendFinder} \footnote{\url{https://github.com/adamkraus/comove}} \citep{Tofflemire2021}. \texttt{FriendFinder} used \textit{Gaia} DR3 positions and parallaxes to identify all sources that fell within our selected three-dimensional search radius around \thisstar. It then calculated the predicted tangential velocity ($v_{tan}$) for every nearby source assuming they have an identical $UVW$ as the source. \texttt{FriendFinder} then compared that value to the true $v_{tan}$, derived from the \textit{Gaia} proper motions. For this grouping, we selected targets with separation $<$25\,pc and a difference in predicted and measured $v_{tan}$ of $<5\,$\kms. A larger physical search radius yielded more objects consistent with membership, but the narrow selection was more than sufficient for aging the star and demonstrating the existence of an association.

\begin{figure}
\centering 
\includegraphics[width=\linewidth]{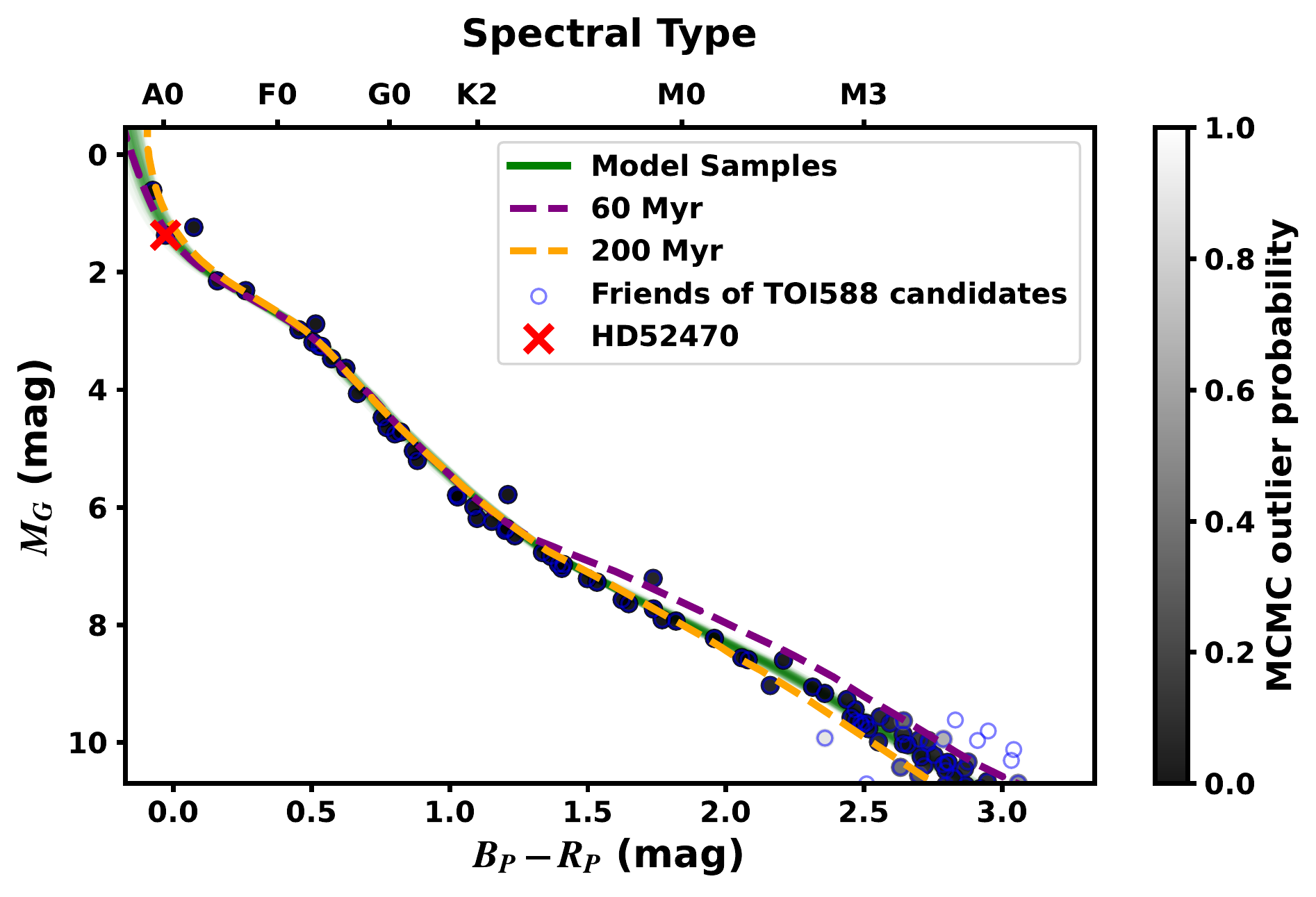}
\caption{Color-magnitude diagram (CMD) of stars spatially and kinematically near \thisstar. The green lines show random draws from our MCMC fit posterior. Points are shaded by the probability that they are part of the outlier model. \thisstar\ is denoted with a red X. The orange and purple lines designate age extremes for reference. }
\label{fig:cmd}
\end{figure}

Our selection yielded 283 stars, including \thisstar. The population color-magnitude diagram (CMD) followed a tight Pleiades-like sequence (Figure~\ref{fig:cmd}). Further, a high fraction of the candidate co-movers had \textit{Gaia} radial velocities consistent with \thisstar. Importantly, \texttt{FriendFinder} did not use radial velocities or CMD information for selection, so consistency here made it clear this is a true co-moving and co-eval population. We denoted this population \association, following the convention from \citet{Tofflemire2021}.

\subsection{Measuring \association's age}

To determine the age of \association\ (and hence the age of \thisstar) we first compared the \textit{Gaia} magnitudes to model isochrones following \citet{Mann2022}. To briefly summarize, we used a mixture model as outlined in \citet{HoggRecipes}. The mixture contained two models, one for the single-star co-eval population, and an outlier population to account for non-members and binaries, inclusion of which tends to bias the isochronal age to older or younger ages, respectively. The fit was done in an MCMC framework using \texttt{emcee} \citep{ForemanMackey:2012}. The six free parameters were age ($\tau$), reddening ($E(B-V)$ [mags]), the amplitude of the outlier population ($P_B$), the offset from the main population to the outlier population ($Y_B$ [mags]), the variance in the outlier population ($V_B$) and a parameter to account for underestimated uncertainties or differential reddening ($f$ [mags]). 

We tested fits using the PARSECv1.2 \citep{PARSEC} and the Dartmouth Stellar Evolution Program \citep[DSEP;][]{Dotter2008} with magnetic enhancement \citep{Feiden2012b}. We initially restricted our analysis to Solar metallicity, but tested [M/H]=-0.1 and +0.1 with the PARSEC models. The DSEP-magnetic models were only available at Solar metallicity and did not extend to the highest-mass stars in the group. We ran the fit with 20 walkers for 10,000 steps following a burn-in of 2,000 steps, which was $>$50 times the autocorrelation time. 

As we show in Figure~\ref{fig:cmd}, ages $\gtrsim200$\,Myr or $<80$\,Myr failed to reproduce the pre-main-sequence M dwarfs. The isochrone fit yielded an age of $106^{+11}_{-8}$\,Myr, with negligible reddening ($E(B-V)<0.05$) and a small outlier population ($P_B=0.15\pm0.05$). The errors on the age are likely underestimated due to our assumptions and limitations of the models. For example, a slightly metal-rich grid ([M/H]=+0.1) gave a similar fit and yielded an older age $118^{+12}_{-8}$\,Myr. Additional adjustments, such as down-weighting the coolest stars, where models struggle to reproduce observations, changed the age at the 10\,Myr level. 

As an additional constraint on age, we also measured rotation periods for candidate members of \association\ from their \tess\ light curves. The rotation sequence provides an age constraint that is (largely) independent of the isochronal measurement, instead relying on the relation between age, color, and rotation period \citep[Gyrochonology;][]{Barnes2003}.

We followed the method outlined in \citet{Barber2022}. To briefly summarize, we generated \tess\ light curves from the FFI cutouts, first creating raw flux light curves from the FFI cutouts centered on each candidate. Then, we generated a Causal Pixel Model (CPM) of the telescope systematics using the \texttt{unpopular} package \citep{Hattori2021} using the ``Similar Brightness'' to generate the model for each star. We subtracted the resulting CPM systematic model from the initial light curves. In total, we extracted usable light curves for 117 targets; the majority of the remaining were too faint or had significant contaminating flux from nearby stars.

We searched every single-sector light curve for each star for rotation periods from $0.1-30$ days using the Lomb-Scargle algorithm \citep{Lomb:1976}. Each identified period was inspected by eye and assigned a quality score following \citet{Rampalli2021}. We retained periods with a score of Q0 or Q1 (105 stars). 

\begin{figure}
\centering 
\includegraphics[width=\linewidth]{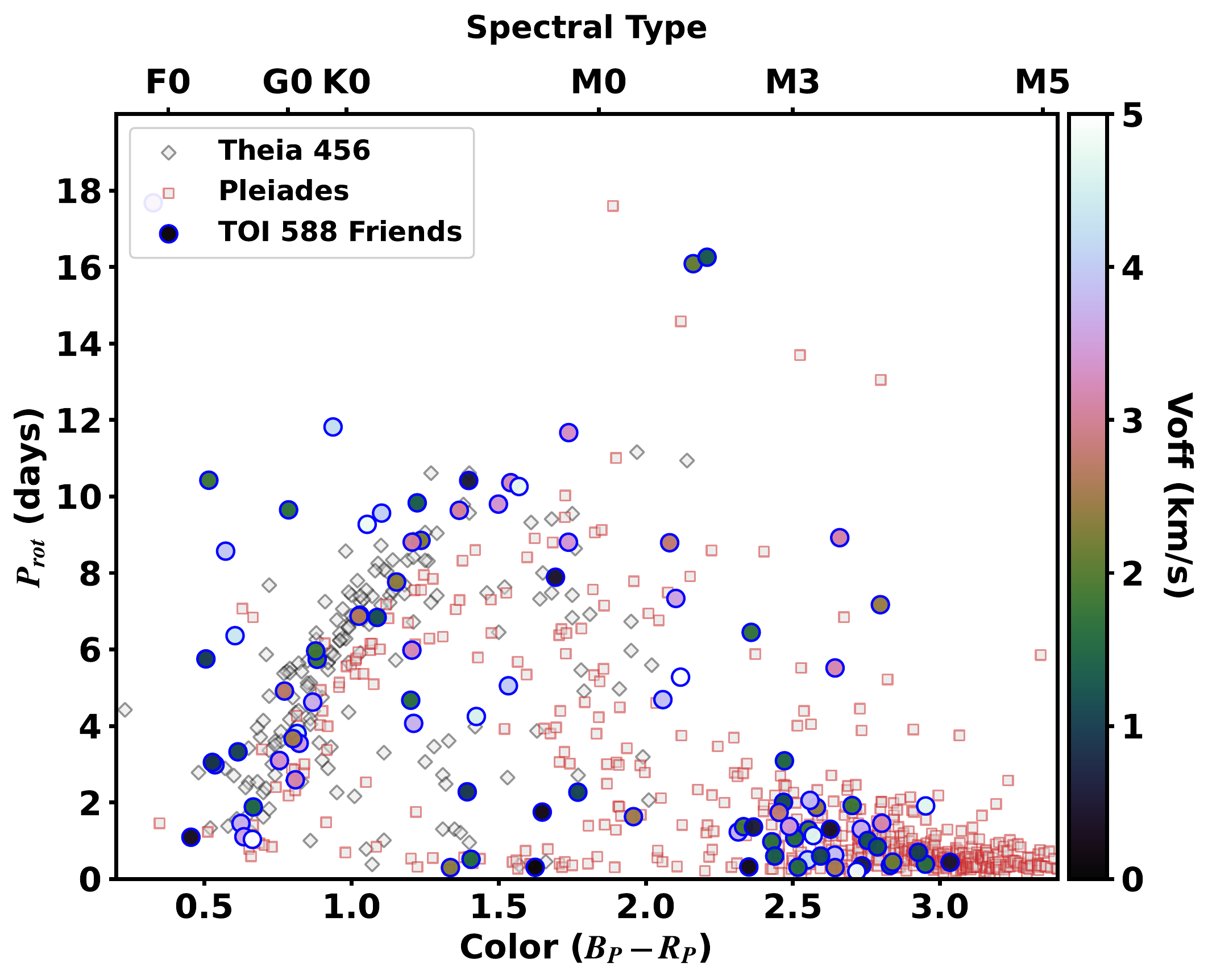}
\caption{Rotation periods of candidate members of \association\ as a function of $B_P-R_P$ color. The rotation periods show a clear sequence as expected for a co-eval association, and the outliers tend to be further from \thisstar\ kinematically (see color-coding). We show the sequences for Theia 456 \citep[150-200\,Myr][]{Andrews2022} and Pleiades \citep[112\,Myr][]{Rebull2016} for comparison. }
\label{fig:prot}
\end{figure}

We show the rotation sequence in Figure~\ref{fig:prot}. The slowly-rotating sequence of FGK dwarfs in \association{} sits above the Pleiades \citep{Rebull2016}, indicating an age $>112$\,Myr \citep{Dahm2015}. The sequence also closely matches that of Theia 456 \citep{Kounkel2019} which was recently determined to be 150-200\,Myr  \citep{Andrews2022}. 

The gyrochronological age is somewhat older than our isochronal value. However, the systematics in the isochronal age were sufficient that the measurements are consistent. We adopted a generous age of 150$\pm$25\,Myr which encompasses both estimates. 

\section{Analysis}
\label{sec:Analysis}
\subsection{EXOFAST\lowercase{v}2 Global Fits}
\label{sec:GlobalModel}
We globally fit all available data using the public exoplanet fitting suite \texttt{EXOFASTv2} \citep{Eastman:2013, Eastman:2019} in order to determine the host star and companion parameters for the \thisstar system (Tables \ref{tab:exofast_planetary} and \ref{tab:TOI588_other}). We fit the Spectral Energy Distribution (SED) and used the MESA Isochrones and Stellar Tracks (MIST) stellar evolution models \citep{Paxton:2011,Paxton:2013,Paxton:2015,Choi:2016,Dotter:2016} in order to constrain the parameters of the host star. We account for smearing from the 30-minute cadence in our FFI light curves from sectors 6 and 8. Our ground-based photometric follow-up from both LCOGT and ASTEP were additively detrended against the parameters described in \S\ref{sec:follow-up} \citep{Collins:2017}.

We initially exclude all ground-based transits from the fit, incorporating only the \tess\ transits, CHIRON radial velocities, and the SED. We turn off the Claret \citep{Claret:2017} tables for these fits since they are less reliable for hot stars (> 10,000 K) and allow the fit to constrain the quadratic limb darkening coefficients directly from the \tess\ transits. The \tess\ 30-minute and 2-minute light curves have out-of-transit standard deviations of 110 ppm and 310 ppm respectively, which was precise enough to independently constrain the limb darkening coefficients in the \tess\ bandpass. We then add in each ground-based transit iteratively, ensuring that each successive fit is still consistent with the \tess\ only fit to 1$\sigma$. Since the ground-based light curves have too low S/N to constrain limb darkening in their respective bandpasses, we apply Gaussian priors on the limb darkening coefficients according to the methods suggested by \citet{Patel2022}. The results presented in Tables \ref{tab:exofast_planetary}, and \ref{tab:TOI588_other} are the final iterations of this process including all ground-based light curves discussed in \S \ref{sec:follow-up}.

We adopted the age, as discussed in \S \ref{sec:ClusterMembership}, of 150$\pm$25\,Myr as a Gaussian prior in our global fits, and placed a Gaussian prior on the parallax from \textit{Gaia} DR3 \citep{Gaia2022} while correcting for the parallax bias according to \citet{Lindegren2021B}. We also placed an upper limit on the line of sight extinction as described in \citet{Schlegel:1998} and \citet{SchlaflyFinkbeiner2011}. Because \thisstar is a rapidly rotating B-star, stellar parameters derived from spectroscopic observations are not reliable. Therefore, we do not obtain precise measurements for the stellar metallicity from our spectral analysis and thus set a Gaussian prior of 0.0 $\pm$ 0.5 dex. We also set a Gaussian prior of 0 $\pm$ 10\% of the contamination ratio reported in the \tess\ Input Catalogue (TIC, \citealp{Stassun:2018_TIC}) in order to fit for a dilution term in the \tess\ band. While the SPOC PDC-SAP light curves are corrected for known nearby companions, fitting for a dilution term accounts for unknown nearby blended stars in the \tess\ aperture and serves as an independent check on the reported contamination correction. We adopted the convergence criteria of \citet{Eastman:2019} which recommend a Gelman-Rubin statistic $<1.01$ and over 1000 independent draws.

\begin{table*}
\scriptsize
\centering
\caption{Median values and 68\% confidence interval for global model of \thisstar}
\begin{tabular}{llcccc}
  \hline
  \hline
\smallskip\\\multicolumn{2}{l}{Stellar Parameters:}&\smallskip\\
~~~~$M_*$\dotfill &Mass (\msun)\dotfill &$2.383^{+0.10}_{-0.095}$\\
~~~~$R_*$\dotfill &Radius (\rsun)\dotfill &$1.863^{+0.087}_{-0.082}$\\
~~~~$F_{Bol}$\dotfill &Bolometric Flux (cgs)\dotfill &$0.0000000492^{+0.000000014}_{-0.0000000097}$\\
~~~~$\rho_*$\dotfill &Density (cgs)\dotfill &$0.519^{+0.076}_{-0.065}$\\
~~~~$\log{g}$\dotfill &Surface gravity (cgs)\dotfill &$4.274^{+0.042}_{-0.040}$\\
~~~~$T_{\rm eff}$\dotfill &Effective Temperature (K)\dotfill &$10400^{+800}_{-660}$\\
~~~~$[{\rm Fe/H}]$\dotfill &Metallicity (dex)\dotfill &$-0.01^{+0.19}_{-0.20}$\\
~~~~$[{\rm Fe/H}]_{0}$\dotfill &Initial Metallicity$^{1}$ \dotfill &$0.02^{+0.18}_{-0.20}$\\
~~~~$Age$\dotfill &Age (Gyr)\dotfill &$0.153\pm0.024$\\
~~~~$EEP$\dotfill &Equal Evolutionary Phase$^{2}$ \dotfill &$314.3^{+7.3}_{-7.4}$\\
~~~~$A_V$\dotfill &V-band extinction (mag)\dotfill &$0.134^{+0.075}_{-0.084}$\\
~~~~$\sigma_{SED}$\dotfill &SED photometry error scaling \dotfill &$1.28^{+0.38}_{-0.26}$\\
~~~~$\varpi$\dotfill &Parallax (mas)\dotfill &$6.483\pm0.049$\\
~~~~$d$\dotfill &Distance (pc)\dotfill &$154.3\pm1.2$\\

\multicolumn{2}{l}{Planetary Parameters:}&b\smallskip\\
~~~~$P$\dotfill &Period (days)\dotfill &$39.471814\pm0.000014$\\
~~~~$R_P$\dotfill &Radius (\rj)\dotfill &$1.580^{+0.074}_{-0.070}$\\
~~~~$M_P$\dotfill &Mass (\mj)\dotfill &$68.0^{+7.4}_{-7.1}$\\
~~~~$T_C$\dotfill &Time of conjunction (\bjdtdb)\dotfill &$2459231.75856\pm0.00021$\\
~~~~$T_0$\dotfill &Optimal conjunction Time$^{3}$ (\bjdtdb)\dotfill &$2458915.98404\pm0.00017$\\
~~~~$a$\dotfill &Semi-major axis (AU)\dotfill &$0.3058^{+0.0042}_{-0.0041}$\\
~~~~$i$\dotfill &Inclination (Degrees)\dotfill &$89.13\pm0.15$\\
~~~~$e$\dotfill &Eccentricity \dotfill &$0.560^{+0.029}_{-0.031}$\\
~~~~$\omega_*$\dotfill &Argument of Periastron (Degrees)\dotfill &$165.9^{+5.5}_{-5.6}$\\
~~~~$T_{eq}$\dotfill &Equilibrium temperature (K)\dotfill &$1237^{+73}_{-61}$\\
~~~~$\tau_{\rm circ}$\dotfill &Tidal circularization timescale$^{4}$ (Gyr)\dotfill &$9800^{+6300}_{-4000}$\\
~~~~$K$\dotfill &RV semi-amplitude (m/s)\dotfill &$2700\pm290$\\
~~~~$R_P/R_*$\dotfill &Radius of planet in stellar radii \dotfill &$0.08715^{+0.00034}_{-0.00036}$\\
~~~~$a/R_*$\dotfill &Semi-major axis in stellar radii \dotfill &$35.3^{+1.6}_{-1.5}$\\
~~~~$\tau$\dotfill &Ingress/egress transit duration (days)\dotfill &$0.02401^{+0.00088}_{-0.00089}$\\
~~~~$T_{14}$\dotfill &Total transit duration (days)\dotfill &$0.26960\pm0.00080$\\
~~~~$b$\dotfill &Transit Impact parameter \dotfill &$0.328^{+0.042}_{-0.053}$\\
~~~~$T_{S,14}$\dotfill &Total eclipse duration$^{5}$ (days)\dotfill &$0.341^{+0.031}_{-0.028}$\\
~~~~$\rho_P$\dotfill &Density (cgs)\dotfill &$21.3^{+4.0}_{-3.4}$\\
~~~~$logg_P$\dotfill &Surface gravity \dotfill &$4.829^{+0.059}_{-0.062}$\\
~~~~$\fave$\dotfill &Incident Flux (\fluxcgs)\dotfill &$0.398^{+0.11}_{-0.075}$\\
~~~~$T_S$\dotfill &Time of eclipse (\bjdtdb)\dotfill &$2459238.47^{+0.75}_{-0.68}$\\
~~~~$e\cos{\omega_*}$\dotfill & \dotfill &$-0.541^{+0.037}_{-0.034}$\\
~~~~$e\sin{\omega_*}$\dotfill & \dotfill &$0.136^{+0.050}_{-0.051}$\\
~~~~$d/R_*$\dotfill &Separation at mid transit \dotfill &$21.3^{+2.3}_{-2.0}$\\
\hline
\hline
\end{tabular}
 \begin{flushleft} 
  \footnotesize{ 
    \textbf{\textsc{\hspace{0.75in}NOTES:}}\\
\hspace{0.75in}See Table 3 in \citet{Eastman:2019} for a list of the derived and fitted parameters in EXOFASTv2.\\
\hspace{0.75in}$^1$Initial metallicity is the metallicity of the star when it formed.\\
\hspace{0.75in}$^2$The Equal Evolutionary Point is a proxy for age and corresponds to static points in a stars evolution when using MIST isochrones. \\
\hspace{0.81in}See \S2 in \citet{Dotter:2016} for a more detailed description of EEP.\\
\hspace{0.75in}$^3$Transit mid-point time that minimizes the covariance between TC and Period. \\
\hspace{0.75in}$^4$The tidal quality factor (Q$_p$) is assumed to be 10$^6$.\\ 
\hspace{0.75in}$^5$All values in this table for the secondary occultation of \thisstar\ b are predicted values from our global analysis.              
               }
 \end{flushleft}
\label{tab:exofast_planetary}
\end{table*}

\begin{table*}
\small
\centering
\caption{Median values and 68\% confidence interval for global model of \thisstar}
\begin{tabular}{llcccccc}
  \hline
  \hline
\smallskip\\\multicolumn{2}{l}{Wavelength Parameters:}&R&z'&TESS\smallskip\\
~~~~$u_{1}$\dotfill &linear limb-darkening coeff \dotfill &$0.043^{+0.061}_{-0.032}$&$0.043^{+0.061}_{-0.032}$&$0.359\pm0.050$\\
~~~~$u_{2}$\dotfill &quadratic limb-darkening coeff \dotfill &$0.032^{+0.078}_{-0.046}$&$0.044^{+0.084}_{-0.054}$&$-0.050^{+0.089}_{-0.086}$\\
~~~~$A_D$\dotfill &Dilution from neighboring stars \dotfill &--&--&$0.0001\pm0.0016$\\
\smallskip\\\multicolumn{2}{l}{Telescope Parameters:}&CHIRON\smallskip\\
~~~~$\gamma_{\rm rel}$\dotfill &Relative RV Offset (m/s)\dotfill &$31130\pm150$\\
~~~~$\sigma_J$\dotfill &RV Jitter (m/s)\dotfill &$0.00^{+400}_{-0.00}$\\
~~~~$\sigma_J^2$\dotfill &RV Jitter Variance \dotfill &$-30000^{+190000}_{-130000}$\\
\smallskip\\\multicolumn{2}{l}{Transit Parameters:}&TESS (30-minute) &TESS (2-minute)\smallskip\\
~~~~$\sigma^{2}$\dotfill &Added Variance \dotfill &$0.0000000045^{+0.0000000022}_{-0.0000000017}$&$0.0000000016^{+0.0000000055}_{-0.0000000051}$&\\
~~~~$F_0$\dotfill &Baseline flux \dotfill &$1.000011\pm0.000014$&$1.000045\pm0.000014$\\

\hline
  \hline
\label{tab:TOI588_other}
 \end{tabular}
\end{table*}

\subsection{Gravity Darkening Fit}
\begin{figure*}
    \centering
    \includegraphics[trim={0cm 0cm 0cm 0cm},clip,scale=0.49]{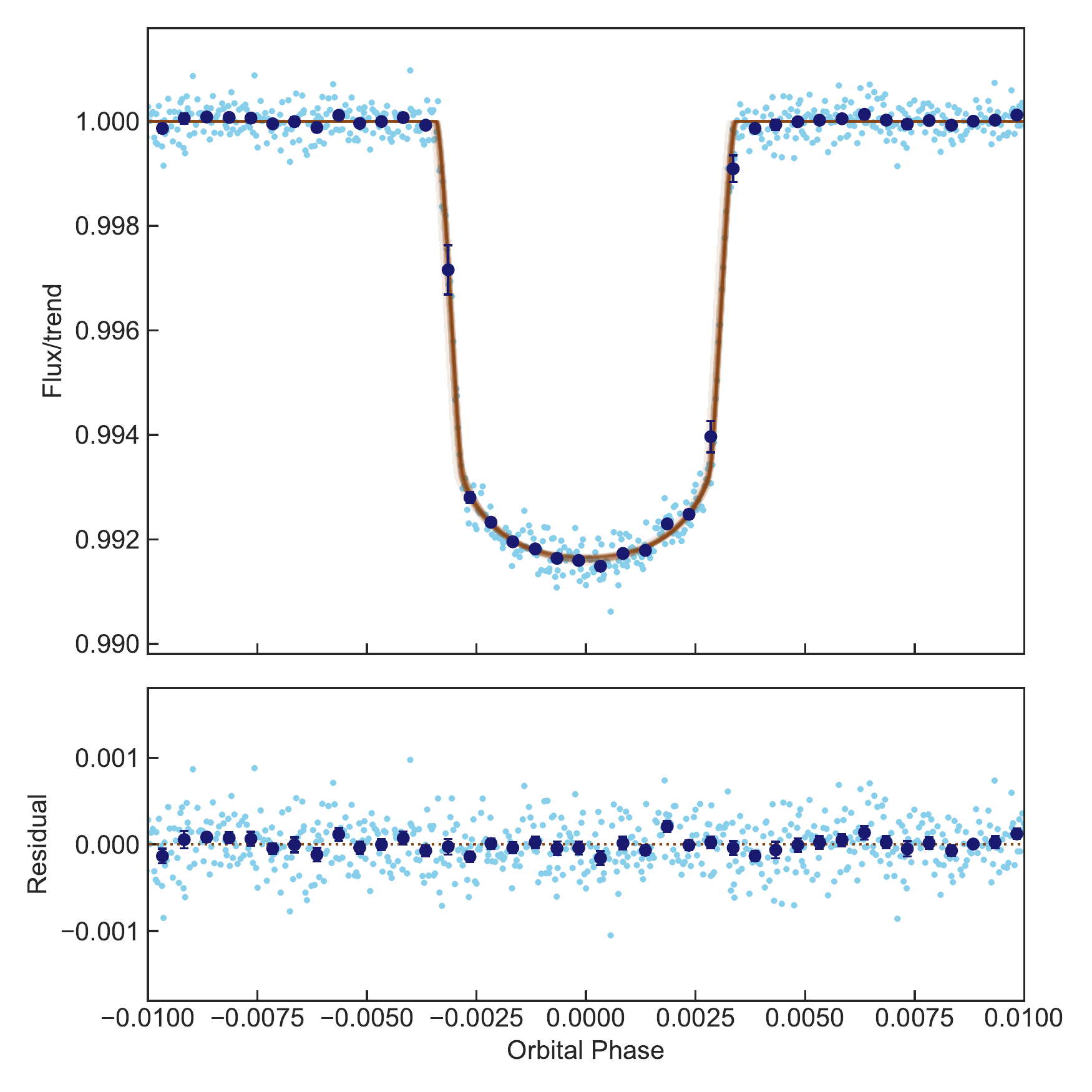}\includegraphics[trim={0cm 0cm 0cm 0cm},clip,scale=0.49]{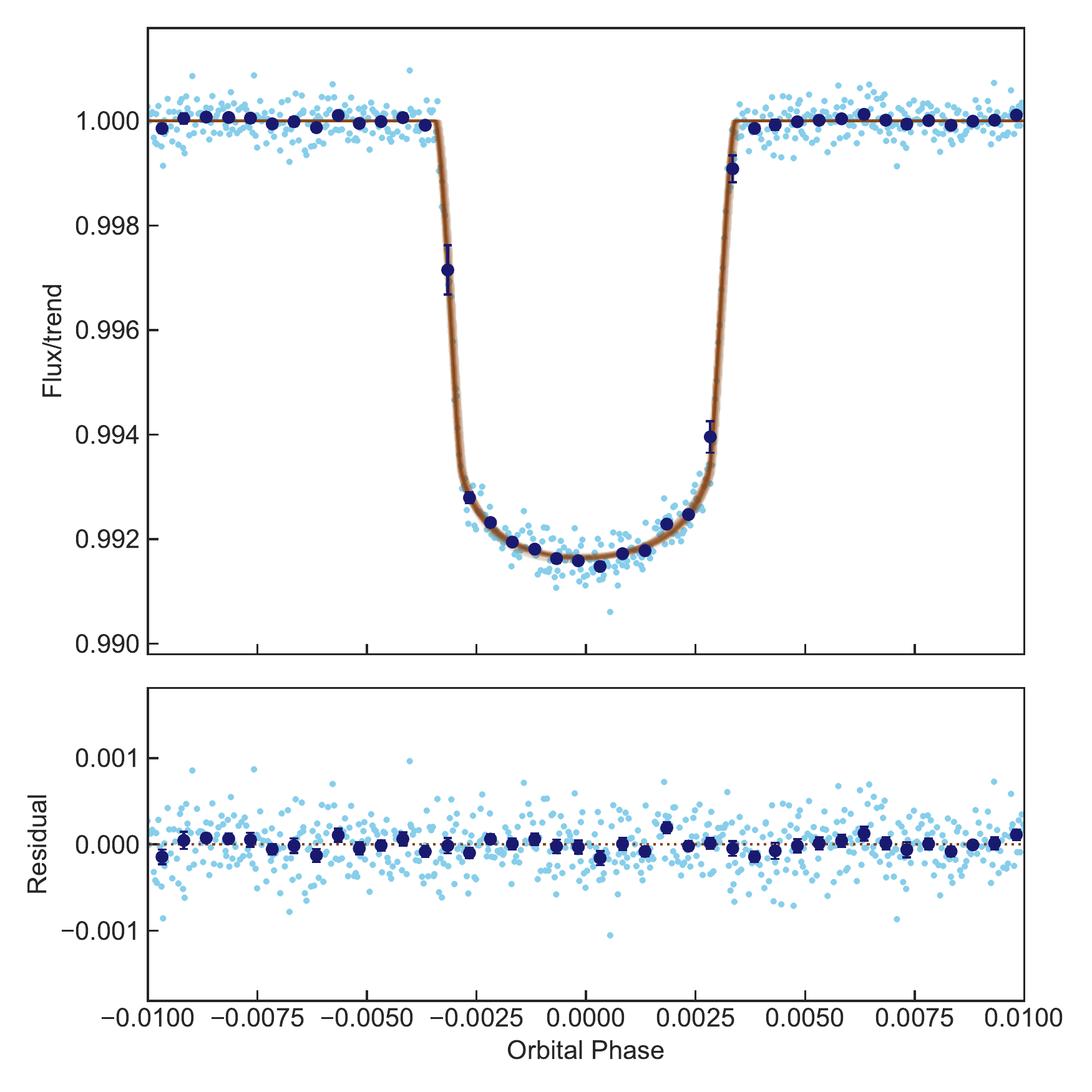}
    \caption{(Left) Plotted are the \tess\ transits phase-folded, binned to 30-minutes, and fit to a symmetric \citet{Mandel:2002} transit model. 
    (Right) The same light curve fit according to \citet{Hooton2022} accounting for asymmetries induced by gravity darkening.}
    \label{fig:GD_figure}
\end{figure*}

Given that \thisstar is a rapidly rotating B-star, we expect that gravity darkening would have a significant effect on the light curve \citep{Barnes2009}. Hence, we performed two additional fits in order to investigate the effects of gravity darkening on \thisstar's light curve. We perform a symmetric fit based on the standard \citet{Mandel:2002} transit model as well as an additional, similar fit following the techniques described in \citet{Hooton2022} in order to account for deviations induced by oblateness and brightness variations that arise from gravity darkening (see Figure \ref{fig:GD_figure}). For this analysis, we fit only the transits from \tess\ because of the much lower signal to noise ratios of the ground-based light curves, as well as the fact that systematics in ground-based observations can imitate the effects of gravity darkening.

In both fits, we adopted Gaussian priors based on the \texttt{EXOFASTv2} outputs where possible. We reparameterized the limb darkening coefficients $u_{1}$, $u_{2}$ taken directly from the \citet{Claret:2017} tables according to \citet{Kipping:2013} and adopted Gaussian priors with standard deviations of 0.5 and 0.1 respectively. We used wide uniform priors on the period, time of conjunction, and planetary radius in addition to a uniform prior ranging from -1 to 1 on $\sqrt{e}\cos{\omega}$ and $\sqrt{e}\sin{\omega}$. We also used a wide uniform prior on the impact parameter \textit{b} for both fits. However, in the case of the gravity darkening fit, we allowed \textit{b} to also sample negative values as we can no longer assume a symmetric stellar disk. Finally, we fixed the gravity darkening exponent $\beta$ according to \citet{Claret2016}.

We found that both of these fits are in good agreement with our global \texttt{EXOFASTv2} fit as all parameters commonly fit among the 3 methods agreed within 2 sigma. We found no significant asymmetries induced by gravity darkening, and therefore we adopt the results from our global \texttt{EXOFASTv2} fit. While our gravity darkening fit is unable to strongly constrain the alignment, the fit favors a potential large misalignment. We encourage additional characterization through Doppler tomography or Rossiter-McLaughlin techniques in order to further constrain the orbital architecture of the \thisstar system.

\subsection{CEPAM Evolutionary Models}
\label{sec:cepam}

\begin{figure}
\centering 
\includegraphics[width=\linewidth]{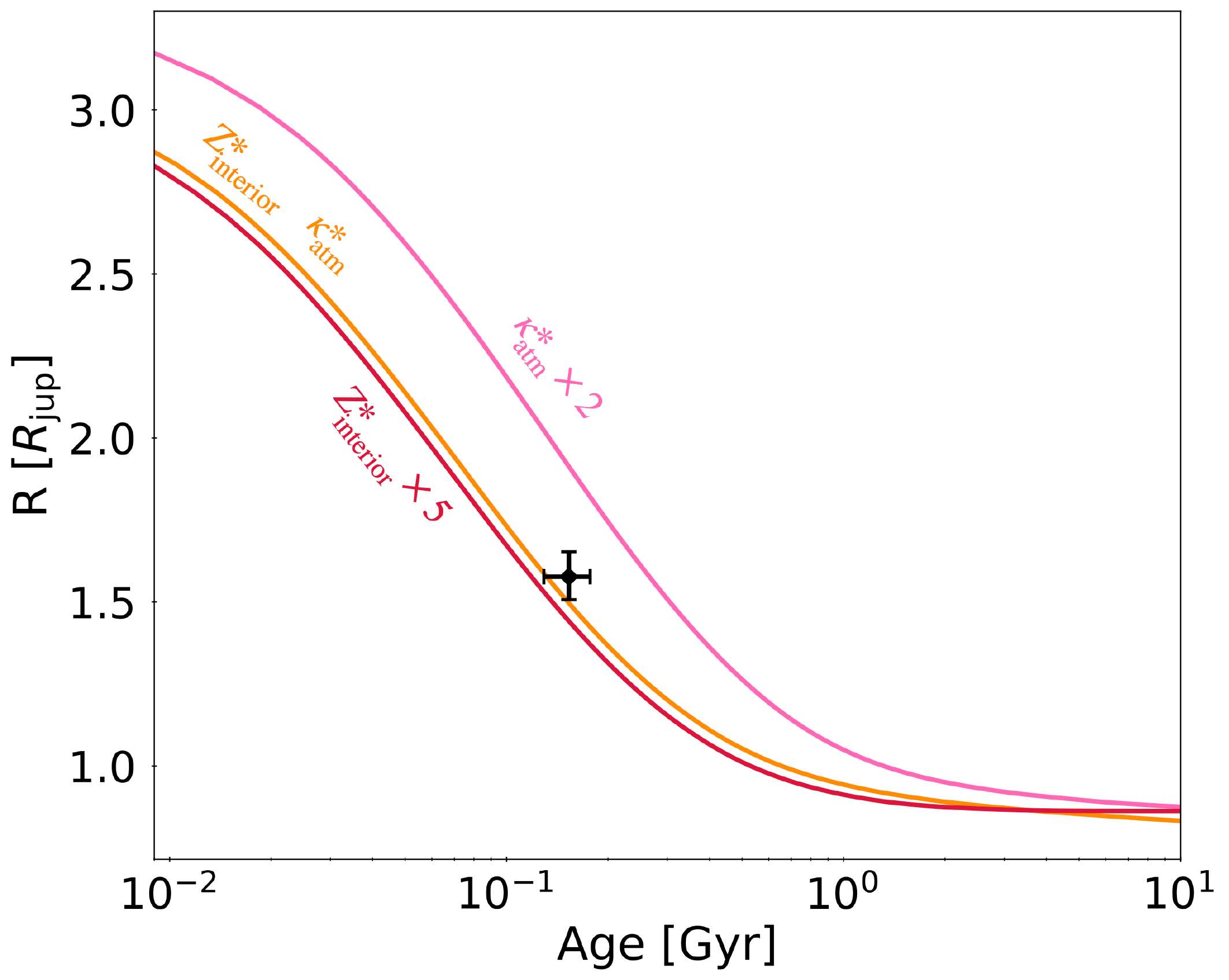}
\caption{The radius evolution of HIP 33609 b. The black point signifies the measured age and radius with 1$\sigma$ error bars. The orange line is our fiducial evolutionary model (see \S \ref{sec:cepam}). The pink line multiplies the atmospheric opacities of the fiducial model by 2, and the red line multiplies the interior metallicity by 5.}
\label{fig:evolution}
\end{figure}

Using CEPAM \citep{Guillot1995}, we calculate evolutionary tracks of \thisplanet. Our models are based on the same approach as in \citet{Bouchy2011}, using the analytical atmospheric boundary conditions from \citet{Guillot2010}. Our fiducial model has a solar metallicity interior ($Z_{\rm interior}^*=Z_{\odot}$) and thermal and visible mean opacities set to $\kappa^*_{\rm th} = 0.04\,\rm g\,cm^{-2}$ and $\kappa^*_{\rm v} = 0.024\,\rm g\,cm^{-2}$, respectively. As shown in Fig. \ref{fig:evolution}, this model reproduces the observed radius for the age of \thisplanet. Because of the brown dwarf’s large mass and intrinsic luminosity ($L_{\rm int}=4\times 10^{30}$  erg/s), we find that the interior is entirely convective and therefore its evolution is not affected by changes of the interior opacities. The energy supplied by tidal dissipation, $L_{\rm tides}\approx 10^{25}\,$erg/s for a tidal quality factor ($Q’=10^6$) (e.g., \citealp{Bodenheimer2001}), is also too low to affect the evolution, as is that due to internal dissipation $L_{\rm dissipation}\approx 10^{27}\,$erg/s (see \citealp{Thorngren2018}).  
 
The radius of \thisplanet is thus mainly affected by three factors: the initial formation entropy (here we assume a hot start initial entropy of $S=13.4$ k$_{B}$/baryon), the deep interior mean molecular weight and the atmospheric opacity (see \citealp{Guillot2005}). Figure \ref{fig:evolution} shows that the latter is by far the dominant effect: when multiplying the atmospheric opacities by 2 over their fiducial values, we obtain a theoretical radius that is 25\% larger (at the measured age) than our fiducial model and clearly incompatible with the observations. On the other hand, when multiplying the interior metallicity by a factor 5 (equivalent to adding $4.2\,M_{\rm Jup}$ of heavy elements in its interior) the radius change remains limited. Although a wider ensemble of dedicated evolution models should be calculated, this already shows that observations of \thisplanet with the James Webb Space Telescope (JWST) would be extremely important, by independently yielding its atmospheric metallicity (that we predict should be solar) and intrinsic luminosity (our evolution models predict $T_{\rm eff}=2630$ K).

\section{Discussion}
\label{sec:results}
\thisplanet joins a population of 37 transiting BDs published to date \citep{Grieves2021, Carmichael2022, Psaridi2022, Sebastian2022}, and is one of the most extreme sub-stellar companions yet discovered. \thisstar not only has a precise age measurement of 150$\pm$25\,Myr, but is also the brightest and hottest host of a transiting BD discovered to date. The unique combination of host star and BD parameters (see Figure \ref{fig:discussion_fig}) make the \thisstar system a benchmark for testing theories of substellar evolution, BD orbital dynamics, and the effects of insolation on BD atmospheres.

The \thisstar system also extends our knowledge of transiting companions around hot stars. All previously discovered transiting companions around B- and A-type stars have orbital periods less than 10 days \citep{Addison2021,Anderson2018,Dorval2020,Gaudi:2017,Giacalone2022,Hellier2019,Johnson2018,Lund:2017,Morton:2016,Shporer2014,Stevens2020, Talens2017,Zhou:2019}, whereas \thisstar provides the first companion on a long period orbit ($\sim$39 days). 

\subsection{Placing \thisstar in Context}
\label{sec:context}
\thisplanet orbits a bright (\textit{V} = 7.3) B-star with a precisely measured age, and joins a growing population of 12 BDs with host stars above the Kraft break ($\sim$6250 K, \citealp{Kraft:1967}), an exciting regime for studying BD evolution in the context of star-like versus planet-like formation. For example, if BDs do indeed follow formation and evolutionary pathways similar to the giant planets, then we can draw comparisons to recent studies focused on hot Jupiters that have observed a discontinuity in stellar obliquity at the Kraft break \citep{Winn_etal2010, Schlaufman2010}. \citet{Rice2022} suggest that this discontinuity may only exist for the low eccentricity population, a trend which would provide strong evidence for high eccentricity migration as the dominant migration mechanism for hot Jupiters. This hypothesis is also supported by the current population of \tess\ discovered giant planets \citep{Rodriguez2022, Yee2022}. If BDs and giant planets undergo similar migratory processes, then they could exhibit the same discontinuity in stellar obliquity. \thisplanet's high eccentricity (\ecc) makes it the second most eccentric BD behind KOI-415 (e = 0.698) \citep{Moutou2013}. However, \thisstar is significantly more accessible to follow-up because it is more than 500 times brighter than KOI-415 (\textit{V} = 14.2). 

\begin{figure*}
\centering 
\includegraphics[width=\linewidth]{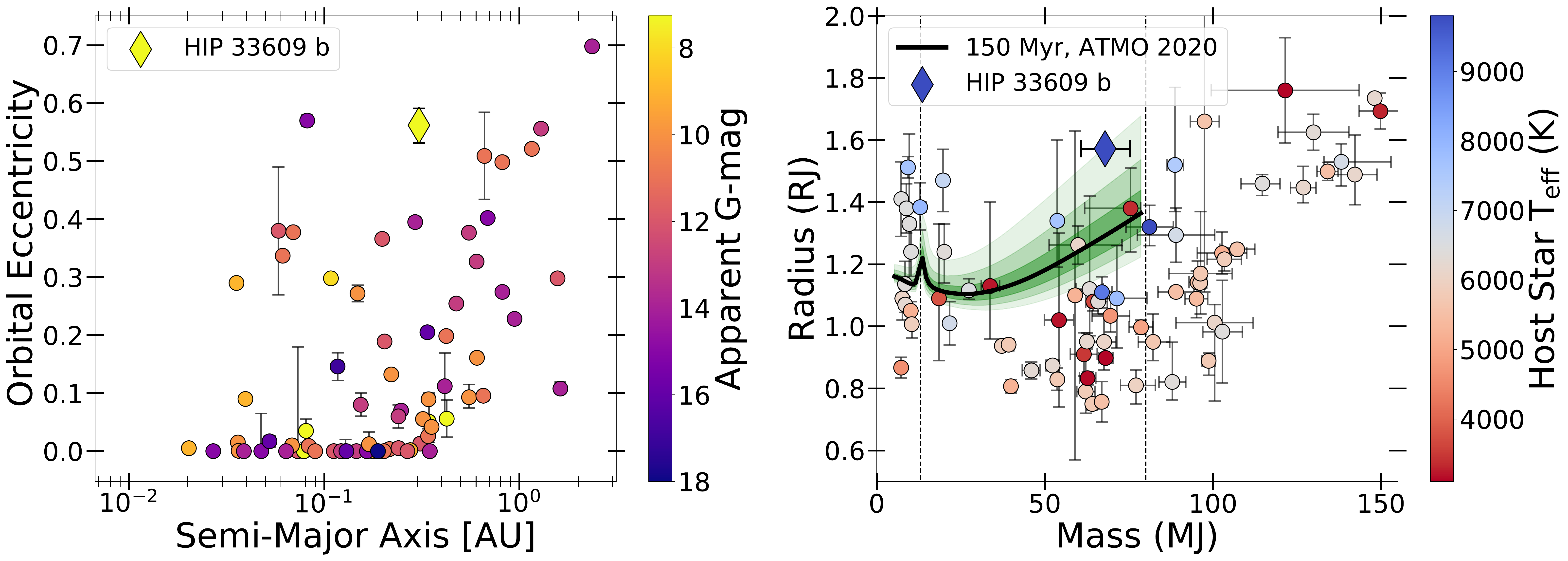}
\caption{(Left) The population of stellar companions ranging from 7-150 $M_J$ in eccentricity and semi-major axis, colored by the apparent magnitude. (Right) The same population in radius and mass, colored by the effective temperature of the host star. Vertical lines at 13 $M_J$ and 80 $M_J$ denote the traditional boundaries of the BD regime. The solid black line shows the ATMO 2020 substellar evolutionary model \citep{Phillips2020} for \thisplanet's measured age (150$\pm$25\,Myr) with the green shaded regions depicting the 1, 2, and 3$\sigma$ uncertainties. \textbf{Note:} systems where the primary body is a BD or white dwarf are not included. \textbf{References:} \citep{Bakos:2010, Buchhave:2011, Tingley2011, Parviainen2014, Bonomo2015, Esteves2015, Stassun2017, Bento2018, Cooke2020, Cortez-Zuleta2020, Grieves2021, Carmichael2022, Gill2022, Psaridi2022, Sebastian2022}}.
\label{fig:discussion_fig}
\end{figure*}

Furthermore, \thisplanet's radius is among the largest for transiting BDs at \rplanet. Substellar evolutionary models predict that BDs form highly inflated and then contract, rapidly at first, then slowing over the course of $\sim$10 Gyr \citep{Baraffe2003, SaumonMarley2008, Burrows2001, Phillips2020}. \thisplanet's large radius is consistent within 3$\sigma$ for substellar models at the estimated age from our analysis using both CEPAM (Figure \ref{fig:evolution}) and the ATMO 2020 models (Figure \ref{fig:discussion_fig}).

\subsection{Future Characterization Prospects}
As the brightest and hottest host star for a transiting BD, the \thisstar system is well-situated for future characterization via ground-based observations. Although the long transit duration ($\sim$6.5 hours) will make spin-orbit alignment measurements challenging, \thisplanet would be a valuable addition to the population of BDs with well-constrained stellar obliquities.

We predict a Rossiter-Mclaughlin semi-amplitude to first order of 270 m/s for \thisplanet using the methods in \citet{Triaud2018}. While this is indeed well below our typical radial velocity uncertainty of $\sim$1000 m/s with CHIRON, we expect orbital obliquity measurements to be more accessible to Doppler Tomographic techniques \citep{Zhou:2016, CollierCameron:2010} given that we can resolve the rotation of the host star. A finding that \thisplanet is highly misaligned, as tentatively indicated by our gravity darkening fit, would provide strong evidence for a dynamically active history. We also encourage future measurements of more BD obliquities in general, in order to provide insight into BD evolutionary pathways similar to that obtained from the obliquity studies of hot Jupiters discussed in \S \ref{sec:context}. 

\thisplanet is also a prime candidate for studying the effects of insolation on BD and giant planet atmospheres. As a long period, massive companion, it can thereby extend the extensive studies of the irradiated atmospheres of BDs and giant planets. The BD KELT-1 b \citep{Siverd:2012, Beatty2014, vonEssen2021}, and giant planets KELT-9 b \citep{Gaudi:2017, Yan2018, Hoeijmakers2018} and TOI-1431 b \citep{Stangret2021} are all ultrahot (T$_{\rm eq}$ $> 2000$ K) companions with well-studied atmospheres that receive approximately 10 times more incident flux than \thisplanet.

\section{Conclusion}
\label{sec:conclusion}
In this paper, we present the discovery of a benchmark transiting BD in the \thisstar system. We use a combination of spectroscopic and photometric observations from both ground- and space-based facilities in order to characterize the host star and transiting BD. \thisstar is a bright (\textit{V} = 7.3), rapidly rotating B-star with an effective temperature of \teffstar. \thisplanet is an inflated BD with a radius of \rplanet\ and a mass of \mplanet\ on a long period (\period), eccentric orbit (\ecc). We also present the discovery of MELANGE-6, a new, young stellar association, of which \thisstar is shown to be a member. We determine the age of the association (and hence \thisstar) to be 150$\pm$25\,Myr. Thus, the \thisstar system is an ideal candidate for testing substellar evolutionary models, as well as for a comparative analysis relative to the extensive population of highly irradiated, short period BDs and giant planets. We encourage the pursuit of stellar obliquity measurements for \thisplanet and the transiting BD population as a whole in order to provide more insight into the formation and evolutionary history of transiting BDs.

\section*{Acknowledgements}
\label{sec:acknowledgements}

A.W.M. was supported by a grant from NASA's Exoplanet research program (80NSSC21K0393). D.R. was supported by NASA under award number NNA16BD14C.

Resources supporting this work were provided by the NASA High-End Computing (HEC) Program through the NASA Advanced Supercomputing (NAS) Division at Ames Research Center for the production of the SPOC data products.

Funding for the TESS mission is provided by NASA's Science Mission Directorate. We acknowledge the use of public TESS data from pipelines at the TESS Science Office and at the TESS Science Processing Operations Center. This research has made use of the Exoplanet Follow-up Observation Program website, which is operated by the California Institute of Technology, under contract with the National Aeronautics and Space Administration under the Exoplanet Exploration Program. This paper includes data collected by the TESS mission that are publicly available from the Mikulski Archive for Space Telescopes (MAST). This research has made use of the NASA Exoplanet Archive, which is operated by the California Institute of Technology, under contract with the National Aeronautics and Space Administration under the Exoplanet Exploration Program.

This work makes use of observations from the LCOGT network. Part of the LCOGT telescope time was granted by NOIRLab through the Mid-Scale Innovations Program (MSIP). MSIP is funded by NSF.

This work makes use of observations from the ASTEP telescope. ASTEP benefited from the support of the French and Italian polar agencies IPEV and PNRA in the framework of the Concordia station program, from OCA, INSU, Idex UCAJEDI (ANR- 15-IDEX-01), the University of Birmingham and ESA.

Some of the observations in the paper made use of the High-Resolution Imaging instrument Zorro obtained under Gemini LLP Proposal Number: GN/S-2021A-LP-105. Zorro was funded by the NASA Exoplanet Exploration Program and built at the NASA Ames Research Center by Steve B. Howell, Nic Scott, Elliott P. Horch, and Emmett Quigley. Zorro was mounted on the Gemini North (and/or South) telescope of the international Gemini Observatory, a program of NSF’s OIR Lab, which is managed by the Association of Universities for Research in Astronomy (AURA) under a cooperative agreement with the National Science Foundation. on behalf of the Gemini partnership: the National Science Foundation (United States), National Research Council (Canada), Agencia Nacional de Investigación y Desarrollo (Chile), Ministerio de Ciencia, Tecnología e Innovación (Argentina), Ministério da Ciência, Tecnologia, Inovações e Comunicações (Brazil), and Korea Astronomy and Space Science Institute (Republic of Korea).

This research received funding from the European Research Council (ERC) under the European Union's Horizon 2020 research and innovation programme (grant agreement n$^\circ$ 803193/BEBOP), and from the Science and Technology Facilities Council (STFC; grant n$^\circ$ ST/S00193X/1).

This work has been carried out within the framework of the NCCR PlanetS supported by the Swiss National Science Foundation under grants 51NF40\_182901 and 51NF40\_205606

\bibliographystyle{apj}

\bibliography{refs}

\end{document}